\documentclass[namedreferences]{solarphysics}
\usepackage[hyperref,optionalrh]{spr-sola-addons}
\usepackage{graphicx}
\usepackage{amssymb}            
\usepackage{color}                                        
\usepackage{breakurl}

\newcommand{\arcsec}{^{\prime\prime}}

\chardef\us=`\_

\begin{document}

\begin{article}
\begin{opening}
 
\title{Transverse Oscillations of Coronal Loops Induced by a Jet-Related Confined Flare on 11 July 2022}

\author[addressref={aff1,aff2}]{\inits{M.S.}\fnm{Musheng}~\lnm{Lin}}
\author[addressref={aff1}]{\inits{Y.}\fnm{Ya}~\lnm{Wang}\orcid{0000-0003-3699-4986}}
\author[addressref={aff3}]{\inits{L.H.}\fnm{Liheng}~\lnm{Yang}\orcid{0000-0003-0236-2243}}
\author[addressref={aff4}]{\inits{J.}\fnm{Jie}~\lnm{Chen}\orcid{0000-0001-7472-5539}}
\author[addressref={aff1,aff2}]{\inits{W.W.}\fnm{Wenwei}~\lnm{Pan}}
\author[addressref={aff5,aff1}]{\inits{S.Y.}\fnm{Shuyue}~\lnm{Li}}
\author[addressref={aff1},corref,email={zhangqm@pmo.ac.cn}]{\inits{Q.M.}\fnm{Qingmin}~\lnm{Zhang}\orcid{0000-0003-4078-2265}}

\address[id=aff1]{Key Laboratory of Dark Matter and Space Astronomy, Purple Mountain Observatory, CAS, Nanjing 210023, China}
\address[id=aff2]{School of Astronomy and Space Science, University of Science and Technology of China, Hefei 230026, China}
\address[id=aff3]{Yunnan Observatories, Chinese Academy of Sciences, Kunming 650216, China}
\address[id=aff4]{National Astronomical Observatories, Chinese Academy of Sciences, Beijing 100012, China}
\address[id=aff5]{School of Science, Nanjing University of Posts and Telecommunications, Nanjing 210023, China}

\runningauthor{M.S. Lin et al.}
\runningtitle{Transverse Oscillations of Coronal Loops}

\begin{abstract}
In this article, we report the multiwavelength and multiview observations of transverse oscillations of two loop strands 
induced by a jet-related, confined flare in active region NOAA 13056 on 11 July 2022.
The jet originates close to the right footpoint of the loops and propagates in the northeast direction.
The average rise time and fall time of the jet are $\approx$ 11 and $\approx$ 13.5 minutes, so that the lifetime of the jet reaches $\approx$ 24.5 minutes.
The rising motion of the jet is divided into two phases with average velocities of $\approx$ 164 and $\approx$ 546\,km\,s$^{-1}$.
The falling motion of the jet is coherent with an average velocity of $\approx$ 124\,km\,s$^{-1}$.
The transverse oscillations of the loops, lasting for 3 $-$ 5 cycles, are of fundamental standing kink mode.
The maximal initial amplitudes of the two strands are $\approx$ 5.8 and $\approx$ 4.9 Mm.
The average periods are $\approx$ 405\,s and $\approx$ 407\,s. Both of the strands experience slow expansions during oscillations.
The lower limits of the kink speed are 895$_{-17}^{+21}$\,km\,s$^{-1}$ for loop\_1 and 891$_{-35}^{+29}$\,km\,s$^{-1}$ for loop\_2, respectively.
The corresponding lower limits of the Alfv\'{e}n speed are estimated to be 664$_{-13}^{+16}$\,km\,s$^{-1}$ and 661$_{-26}^{+22}$\,km\,s$^{-1}$.
\end{abstract}
\end{opening}

\section{Introduction}  \label{intro}
Solar jets are transient and collimated plasma ejections along straight or slightly twisted magnetic field lines, including spicules \citep{bec72,dp07,san19}, 
H$\alpha$ surges \citep{roy73,chae99,jia07}, chromospheric jets \citep{shi07,liu11,sin12,tian14b,wy23}, 
and coronal jets \citep{shi96,cir07,chen12,zqm14a,chen15,chen17,ste15,yang19,duan24,ylh24}.
Most of those jets are generated by impulsive releases of magnetic free energy via magnetic reconnection \citep{yoko96,mo08,nis08,par09,zqm12,mu16,pan16,mar17,wp18,nob22}.
Coronal jets were discovered by the Soft X-ray Telescope (SXT) on board the Yohkoh spacecraft \citep{shi92}.
They are located at the boundaries of active regions (ARs) or in coronal holes and are regularly observed 
in soft X-ray (SXR) and extreme ultraviolet (EUV) wavelengths \citep[see reviews][and references therein]{Raouafi2016, Shen2021}.
According to the morphology, coronal jets are divided into the anemone type and two-sided type \citep{shi94,shen19}.
Considering that a great number of jets results from eruptions of filaments or minifilaments \citep{hong16,ste16,ylp24},
they could also be classified into standard jets and blowout jets \citep{mo10,pu13,ste22}.
\citet{mo13} investigated 54 polar jets observed simultaneously in SXR and 304 {\AA}.
It is found that a cool ($T\sim10^5$ K) component is present in nearly all blowout jets and in a small minority of standard jets.
Moreover, the spire widths of blowout jets are larger than those of standard jets \citep{ste22}.
High-resolution observations reveal the existence of tiny and recurrent plasmoids in jets, which are explained by the tearing-mode instability in a current sheet 
near the jet base \citep{zqm14b,ni17,jos18,chen22,man22a,hou24}.

\citet{shi00} studied the physical properties of 16 SXR jets observed by Yohkoh/SXT, including the temperature, density, thermal energy, and apparent speed.
They concluded that SXR jets are evaporation flows produced by magnetic reconnection heating.
\citet{nis09} investigated the properties of polar EUV jets 
observed by the Extreme UltraViolet Imager (EUVI) of the Sun-Earth Connection Coronal and Heliospheric Investigation \citep[SECCHI;][]{how08} 
on board the Solar TErrestrial RElations Observatory \citep[STEREO;][]{kai08} ahead (hereafter STA) and behind (hereafter STB) satellites. 
The typical lifetimes are 20 $-$ 30\,minutes. The average velocities in 171 {\AA}  and 304 {\AA} are 400 and 270\,km\,s$^{-1}$, respectively. 
In a further study of AR jets with the Atmospheric Imaging Assembly \citep[AIA;][]{lem12} on board the Solar Dynamics Observatory (SDO), 
\citet{mu16} found that the lifetimes range from 5 to 39\,minutes with a mean value of 18\,minutes and the speeds range from 87 to 532\,km\,s$^{-1}$ with a mean value of 271\,km\,s$^{-1}$. 
Besides, all the jets in their study are co-temporally associated with H$\alpha$ surges.

Magnetohydrodynamic (MHD) waves and oscillations are prevalent in the solar atmosphere \citep[][and references therein]{naka21,wang21,zim21}.
Kink oscillations of coronal loops induced by the flare on 14 July 1998 were first detected by the Transition Region And Coronal Explorer \citep[TRACE;][]{han99} mission.
Magnetic field strengths of the oscillating loops are estimated based on coronal seismology \citep{asch99,naka99,naka01}.
Recently, using the high-resolution observations with the Upgraded Coronal Multi-channel Polarimeter, \citet{yzh24} derived 114 magnetograms of the global corona above the solar limb.
The polarization of kink oscillations could be horizontal \citep{asch02,ver09,asch11,wht12a,nis13,nis17,shi22} or vertical \citep{wang04,gos12,wht12b,sri13,kim14}. 
The length of coronal loops, initial displacement amplitude, period, and damping time of kink oscillations 
lie in the ranges of 78 $-$ 532\,Mm, 0.6 $-$ 31.8\,Mm, 2.07 $-$ 28.19\,minutes, and 2.69 $-$ 35.01\,minutes, respectively. 
The damping time is roughly proportional to the period \citep{ver13, god16a}.
The quality factor ($q=\frac{\tau}{P}$) of kink oscillations is inversely proportional to the square root of the oscillation amplitude \citep{god16b},
where $P$ and $\tau$ represent the period and damping time.
Apart from damping oscillations, non-damping or decayless kink oscillations with smaller amplitudes are found to be important in coronal heating \citep{tian12,anf15,zqm20,gao22,man22b,lid23,zhong23}.

Horizontal oscillations of coronal loops are induced by flare-induced blast waves \citep{naka99}, lower coronal eruptions/ejections \citep[LCEs;][]{zim15}, and EUV waves \citep{shen12,kum13}.
For vertical oscillations of coronal loops, the ways of excitation are diverse, such as magnetic implosion during flares \citep{sim13}, reconnection outflows from flare current sheets \citep{ree20},
filament eruptions \citep{mro11,zqm22a}, EUV waves \citep{zqm22b,zqm23}, and coronal rains \citep{koh17,ver17}.
So far, transverse oscillations of coronal loops excited by coronal jets have rarely been observed and reported.
\citet{sar16} investigated transverse oscillations in a coronal loop, which are triggered by a coronal jet originating from a region close to the loop on 19 September 2014.
Using the loop length (377 $-$ 539\,Mm) and period of oscillation ($\approx$ 32\,minutes), the magnetic field inside the oscillating loop is estimated to be 2.7 $-$ 4.5\,G.
\citet{dai21} studied the transverse oscillation of a coronal loop, which is induced by a blowout jet associated with a C4.2 flare on 16 October 2015.
The initial amplitude, average period, and damping time are $\approx$ 13.6\,Mm, $\approx$ 462\,s, and $\approx$ 976\,s, respectively. 
The magnetic field inside the loop is estimated to be 30 $-$ 43\,G using coronal seismology.
On 11 July 2022, a jet occurred around 01:10 UT, which was accompanied by a C3.5 confined flare in NOAA AR 13056 (S16E63).
Transverse oscillations of the overlying coronal loops were induced by the jet-related flare.
In this work, we aim to investigate the jet and oscillations using multiwavelength and multiview observations.
The paper is organized as follows. The data analysis is described in Section~\ref{data}. The results are presented in Section~\ref{results}.
Comparisons with previous works are discussed in Section~\ref{discussion}, and a brief summary is given in Section~\ref{summary}.

\begin{figure} 
\centerline{\includegraphics[width=0.6\textwidth,clip=]{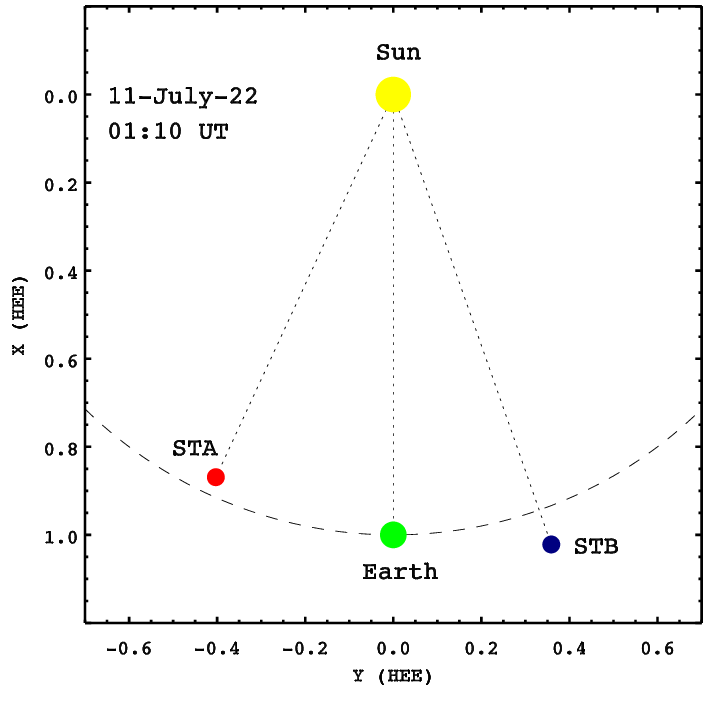}}
\caption{Positions of Earth (green circle), STA (red circle), and STB (blue circle) at 01:10 UT on 11 July 2022.}
\label{fig1}
\end{figure}

\section{Data Analysis} \label{data}
The coronal jet was completely detected by a fleet of ground-based and spaceborne instruments, 
including the Global Oscillation Network Group \citep[GONG;][]{har96} in H$\alpha$ line center, STA/EUVI in 195\,{\AA}, and SDO/AIA in 171, 193, and 304\,{\AA}.
Kink oscillations of the overlying coronal loops were mainly detected in EUV wavelengths.
Full-disk line-of-sight (LOS) magnetograms of the photosphere were observed by the Helioseismic and Magnetic Imager \citep[HMI;][]{sch12} on board SDO. 
SXR fluxes of the C3.5 flare were recorded by the Geostationary Operational Environmental Satellite \citep[GOES;][]{gar94} spacecraft.
Figure~\ref{fig1} shows the positions of Earth (green circle), STA (red circle), and STB (blue circle) at 01:10 UT on 11 July 2022.
STA had a separation angle of 24.9$^{\circ}$ with the Sun-Earth line, while STB did not work.

The level\_1 data of AIA and HMI were calibrated using the standard routines \textsf{aia\_prep.pro} and \textsf{hmi\_prep.pro} built in the Solar Software (SSW). 
The AIA 304\,{\AA} images were coaligned with GONG H$\alpha$ images using the cross-correlation method.
Calibration of the EUVI data was performed using the standard routine \textsf{secchi\_prep.pro}. 
Observational parameters of the instruments are listed in Table~\ref{tab-1}.

\begin{table}
\caption{Description of the observational parameters.}
\label{tab-1}
\tabcolsep 1.5mm
\begin{tabular}{lcccc}
\hline
Instrument& Wavelength   & Cadence   & Pixel Size \\ 
          & [{\AA}]      &  [s]      & [$\arcsec$] \\
\hline
SDO/AIA       & 171, 193, 304 &   12      & 0.6 \\
SDO/HMI       & 6173         &   45      & 0.6 \\
STA/EUVI      & 195          &  150      & 1.6 \\
GONG      & 6562.8       &   60      &  1.1 \\
GOES      & 0.5$-$4   &  2.05     & ... \\
GOES      & 1$-$8     &  2.05     & ... \\
  \hline
\end{tabular}
\end{table}

\section{Results} \label{results}
\begin{figure} 
\centerline{\includegraphics[width=0.9\textwidth,clip=]{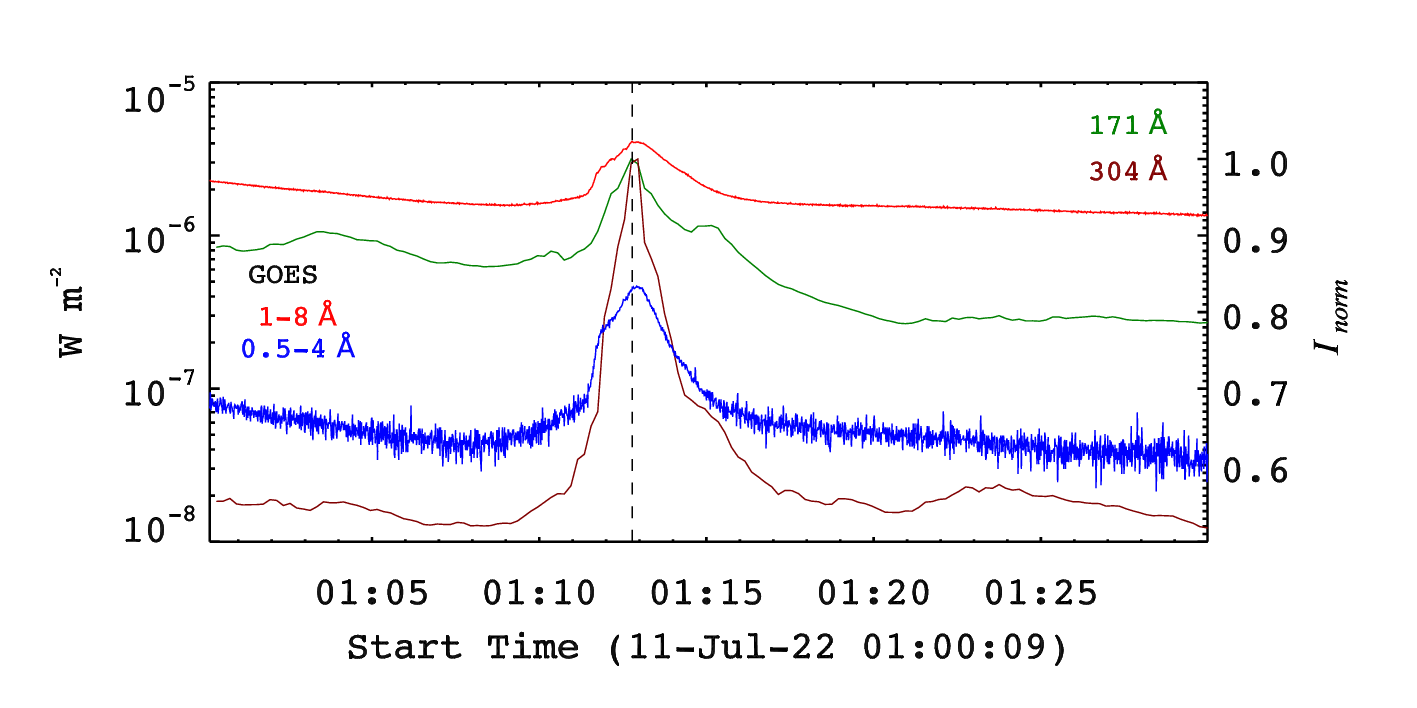}}
\caption{Light curves of the C3.5 flare in 1$-$8\,{\AA} (\textit{red line}), 0.5$-$4\,{\AA} (\textit{blue line}), 171\,{\AA} (\textit{green line}), and 304\,{\AA} (\textit{maroon line}).
The black dashed line denotes the flare peak time at 01:12:47 UT.}
\label{fig2}
\end{figure}

\begin{figure}
\centerline{\includegraphics[width=0.9\textwidth,clip=]{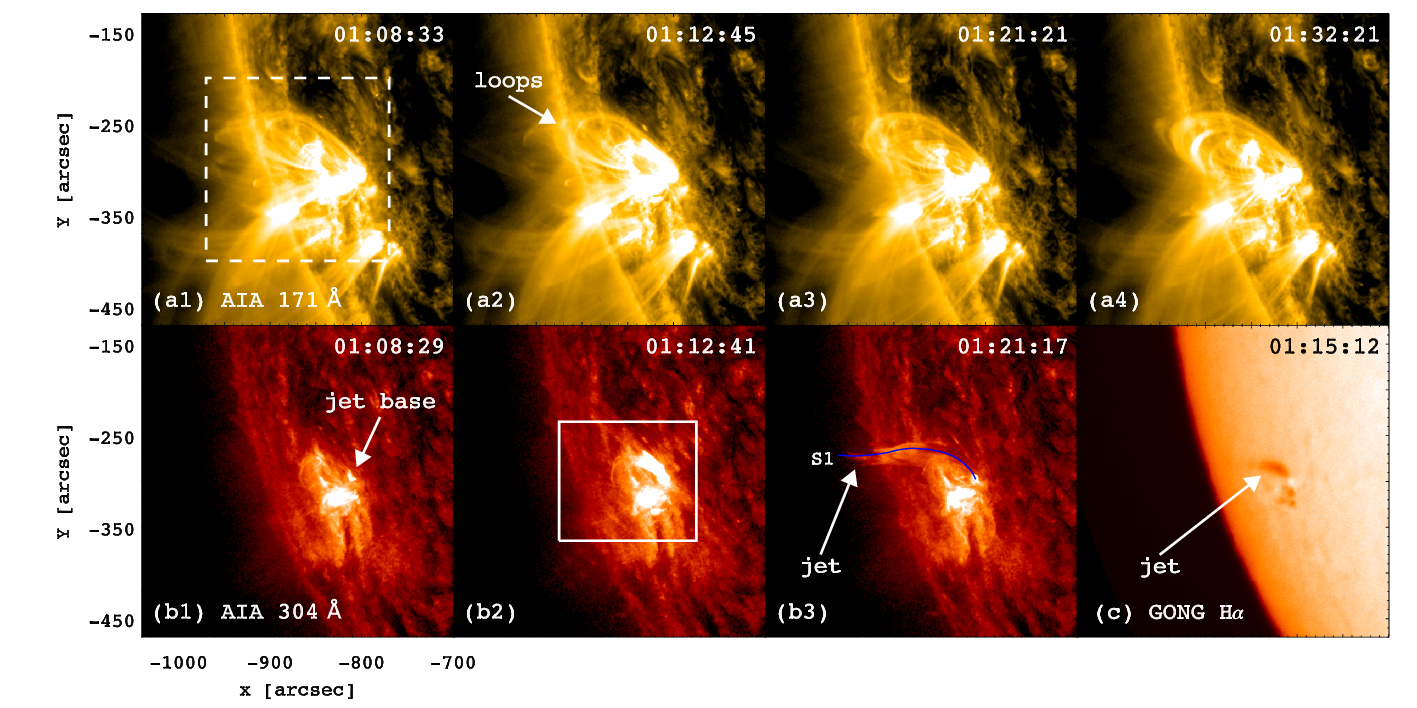}}
\caption{Snapshots of AIA 171\,{\AA} images (a1-a4), 304\,{\AA} images (b1-b3), and an H$\alpha$ image from GONG (c).
The white arrows point to the coronal loops, jet spire, and jet base.
In panel a1, the dashed box represents the field of view (FOV) of Figure~\ref{fig5}a.
In panel b2, the solid box represents the flare region to calculate EUV light curves in 171 and 304\,{\AA}.
In panel b3, a curved slice (S1) is used to investigate the jet evolution.
Animations of 171 and 304\,{\AA} are available in the electronic supplementary material (\textsf{anim1.mp4}).}
\label{fig3}
\end{figure}

In Figure~\ref{fig2}, the blue and red lines show SXR light curves of the C3.5 flare in 0.5$-$4\,{\AA} and 1$-$8\,{\AA}, respectively. 
The short-lived flare starts at $\approx$ 01:08 UT, peaks at $\approx$ 01:12 UT (black dashed line), and ends at $\approx$ 01:16 UT.
Hence, the lifetime of the flare is less than 10\,minutes, which is similar to the jet-related, C1.6 class flare on 15 October 2011 \citep{zqm14a}.
In Figure~\ref{fig3}, the top and bottom panels show the evolutions of the flare and jet in 171 and 304\,{\AA} 
(see online movie \textsf{anim1.mp4} in the electronic supplementary material).
The left panels (a1-b1) show the jet base in AR 13056 at the very beginning of flare. The second column (a2-b2) shows the flare at its maximum with greatly enhanced intensities.
In panel b2, a white box (150$\arcsec\times$130$\arcsec$) is used to calculate the integrated intensities of the flare region. 
The normalized light curves in 171 and 304 {\AA} are plotted with green and maroon lines in Figure~\ref{fig2}, respectively.
It is obvious that EUV emissions of the flare have the same trend and peak time as in SXR.
The third column of Figure~\ref{fig3} shows the jet propagating along curved field lines in the northeast direction. 
The jet appears at $\approx$ 01:10 UT, rises up until $\approx$ 01:21 UT, and falls down along the field lines.
Figure~\ref{fig3}c shows the jet (surge) observed in H$\alpha$ line center at 01:15:12 UT, which has a similar morphology as in 304\,{\AA}. 

\begin{figure} 
\centerline{\includegraphics[width=0.9\textwidth,clip=]{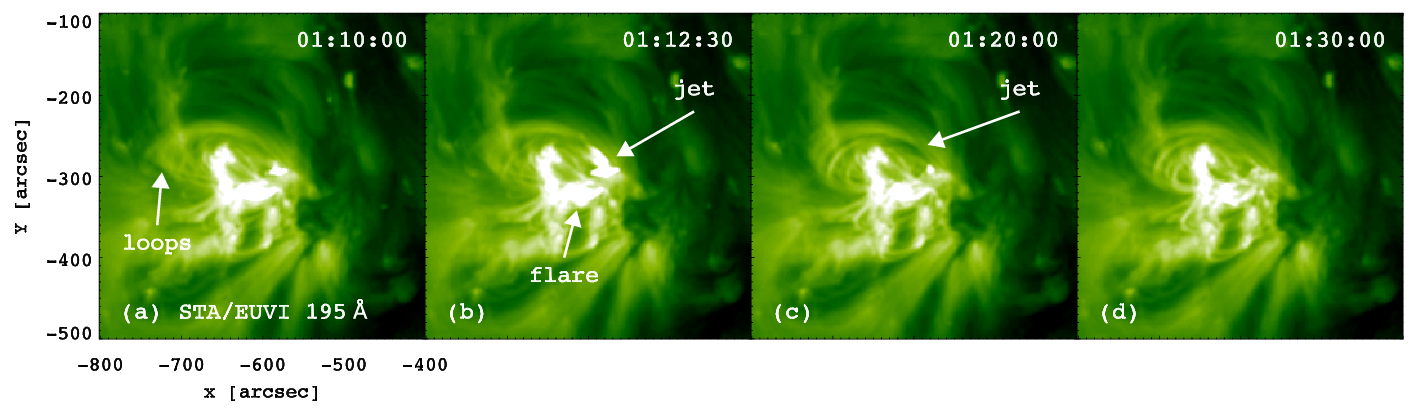}}
\caption{Snapshots of STA/EUVI 195\,{\AA} images. 
The arrows point to the coronal loops at the beginning of eruption (panel a), flare and hot component of the jet (panel b), and cool component of the jet (panel c).
An animation of this figure is available in the electronic supplementary material (\textsf{anim2.mp4}).}
\label{fig4}
\end{figure}

\begin{figure} 
\centerline{\includegraphics[width=0.9\textwidth,clip=]{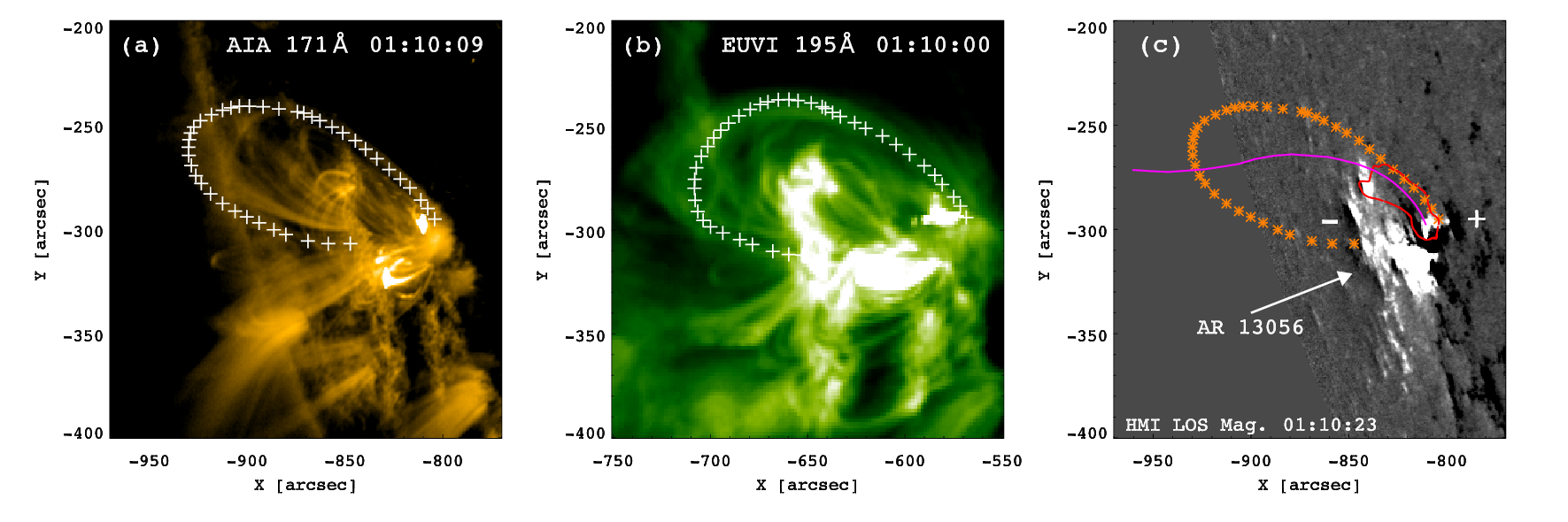}}
\caption{The coronal loops observed by AIA 171\,{\AA} (a) and EUVI 195\,{\AA} (b) at 01:10 UT. 
White ``+" symbols outline the oscillating loop.
(c) HMI LOS magnetogram of AR 13056 at 01:10:23 UT. The orange stars outline the same loop as in panel a.
The red line stands for the intensity contour of the jet observed by AIA 304\,{\AA} at 01:13:05 UT.
The purple line denotes S1 in Figure~\ref{fig3}b3.}
\label{fig5}
\end{figure}

The whole event is also observed by STA/EUVI from a different viewing angle.
Figure~\ref{fig4} shows four snapshots of EUVI 195\,{\AA} images (see online movie \textsf{anim2.mp4} in the electronic supplementary material).
In panel a, the arrow points to the same loops as in AIA 171\,{\AA} at the beginning of eruption.
In panels b and c, the arrows point to the flare, hot component of the jet, and cool component of the jet.
Close-ups of AR 13056 in AIA 171\,{\AA}, EUVI 195\,{\AA}, and HMI LOS magnetogram are displayed in Figure~\ref{fig5}.
In panels a and b, the white ``+" symbols outline the coronal loops at 01:10 UT.
In panel c, the red line stands for the intensity contour of the jet observed by AIA 304\,{\AA} at 01:13:05 UT. 
The orange stars denote the same loop as seen in 171\,{\AA} in panel a. The left and right footpoints are rooted in negative and positive polarities, respectively.
It is clear that the jet base is close to the right footpoint of the coronal loops.

\begin{figure} 
\centerline{\includegraphics[width=0.7\textwidth,clip=]{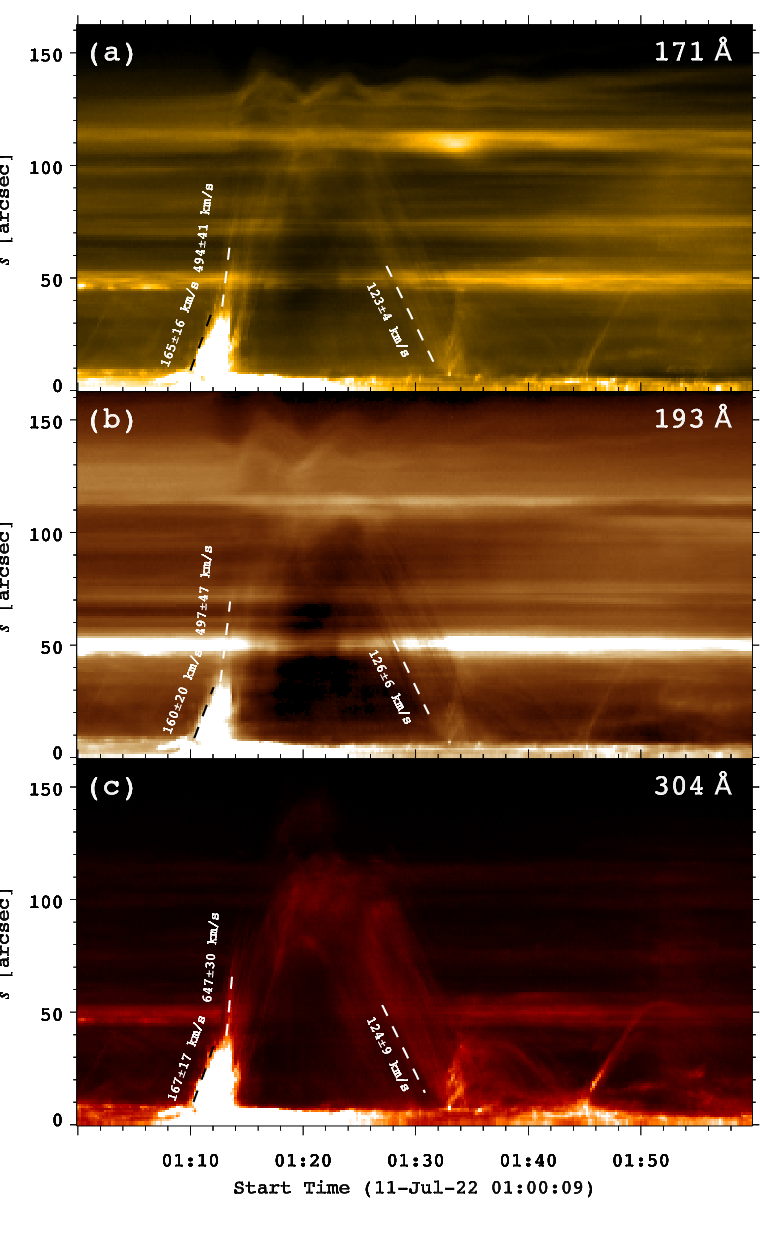}}
\caption{Time-distance diagrams of S1 in AIA 171, 193, and 304\,{\AA}. 
$s=0$ and $s=163\arcsec$ denote the west and east endpoints of S1. The apparent rising and falling speeds of the jet are labeled.}
\label{fig6}
\end{figure}

In order to investigate the kinetic evolution of the jet, a curved slice (S1) is selected along the jet axis in Figure~\ref{fig3}b3.
The total length and width of S1 are 163$\arcsec$ and 3$\arcsec$, respectively.
In Figure~\ref{fig5}c, S1 is overlaid on the magnetogram with a purple line, indicating that the top of the jet apparently reaches and collides with the loops.
Time-distance diagrams of S1 in 171, 193, and 304\,{\AA} are displayed in Figure~\ref{fig6}.
The jet spire demonstrates an asymmetric, parabolic trajectory, which is characterized by a faster rising motion and a slower falling motion \citep{hua20}.
The rising motion is apparently divided into two phases. The first phase is between 01:10:00 UT and 01:12:30 UT. The rising velocity is between 160 and 167\,km\,s$^{-1}$.
The second phase is between 01:12:30 UT and 01:14:00 UT. The rising velocity is between 494 and 647\,km\,s$^{-1}$.
The turning point between the two phases is consistent with the flare peak, implying that the jet may be accelerated by the flare reconnection.

\citet{mo10} drew a schematic picture to illustrate the topology, eruption, and reconnection of a blowout jet (see their Fig. 10).
Two-step magnetic reconnections are involved. The first step is breakout reconnection as the highly sheared core field (filament) starts to rise.
The second step is reconnection beneath the sheared core field (filament) during the blowout eruption.
Similarly, in the schematic cartoon of a minifilament-eruption process, magnetic reconnections take place above and below the minifilament \citep{ste15}.
The cool material opens up through breakout reconnection. The jet becomes more vigorous and propagates along the open field lines.
\citet{wp18} proposed a breakout model for coronal jets with filaments. In their model, 
a filament channel forms beneath a 3D null point as a result of continuous shearing motions.
Meanwhile, a breakout current sheet (BCS) builds up near the null point. As the filament supported by a flux rope erupts, the BCS is strongly squeezed and ramps up.
Below the filament, a flare current sheet (FCS) grows up where magnetic reconnection takes place. After the filament (flux rope) opens up from the BCS, 
it creates an impulsive jet. Therefore, the kinetic evolution of the jet is divided into two phases before and after the flux rope opens up. 
The velocity and kinetic energy of the jet are much larger in the second phase than those in the first phase. 
Such a model is supported by multiwavelength observations of a flare-related, breakout jet on 16 October 2015 \citep{zqm21}.
In the current study, the rising motion of the flare-related jet features a two-step evolution, implying that the jet results from a minifilament eruption.

The falling motion of the jet is coherent, with a velocity of 123$-$126\,km\,s$^{-1}$.
The physical parameters, including the apparent rising speed ($v_r$), falling speed ($v_f$), the velocity ratio $\frac{v_{r}}{v_{f}}$, starting time, and ending time, are listed in Table~\ref{tab-2}.
The average $v_r$ is found to be 1.3 and 4.4 times higher than that of $v_f$.
Moreover, the average rise time ($\approx$ 11\,minutes) is shorter than that of fall time ($\approx$ 13.5\,minutes).
Consequently, the lifetime of jet is $\approx$ 24.5\,minutes, which is nearly half of the jet lifetime on 15 October 2011 \citep{zqm14a}.

\begin{table}
\caption{Physical parameters of the jet in various passbands,
where $v_r$ and $v_f$ stand for the apparent rising and falling speeds.}
\label{tab-2}
\tabcolsep 1.5mm
\begin{tabular}{ccccccc}
\hline
$\lambda$ & $v_r$ & $v_f$ & $v_{r}/v_{f}$ & Start Time & End time & Lifetime \\  
{[{\AA}]} & [km\,s$^{-1}$] & [km\,s$^{-1}$] & - & [UT] & [UT] & [Minute] \\
\hline
171  & 165$\pm$16, 494$\pm$41 &  123$\pm$4 & 1.3, 4.0 & 01:10:00 & 01:33:51 & $\approx$ 24.0 \\
193  & 160$\pm$20, 497$\pm$47 &  126$\pm$6 & 1.3, 3.9 & 01:10:00 & 01:34:25 & $\approx$ 24.5 \\
304  & 167$\pm$17, 647$\pm$30 &  124$\pm$9 & 1.3, 5.2 & 01:10:00 & 01:34:55 & $\approx$ 25.0 \\
\hline
Avg. & 164, 546 & 124 & 1.3, 4.4 & 01:10:00 & 01:34:25 & $\approx$ 24.5 \\
\hline
\end{tabular}
\end{table}

\begin{figure} 
\centerline{\includegraphics[width=0.7\textwidth,clip=]{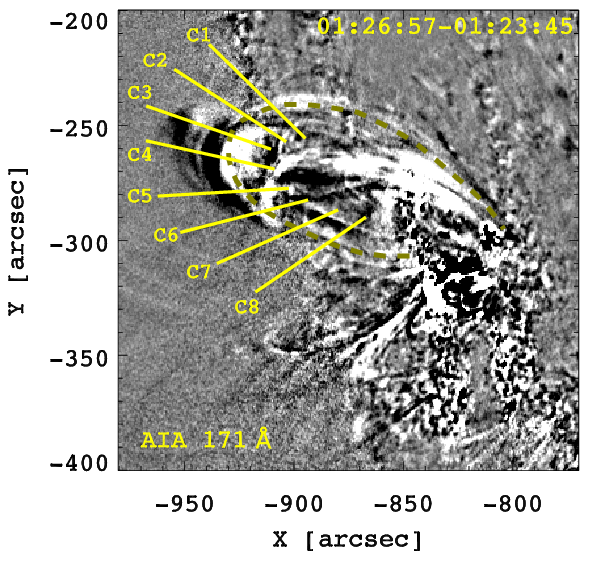}}
\caption{AIA difference image in 171 {\AA} produced by subtracting the image taken at 01:26:57 UT from the one at 01:23:45 UT. 
Eight slices (C1$-$C8), which are 60$\arcsec$ in length, are selected to investigate kink oscillations of the coronal loops.}
\label{fig7}
\end{figure}

\begin{figure} 
\centerline{\includegraphics[width=1.0\textwidth,clip=]{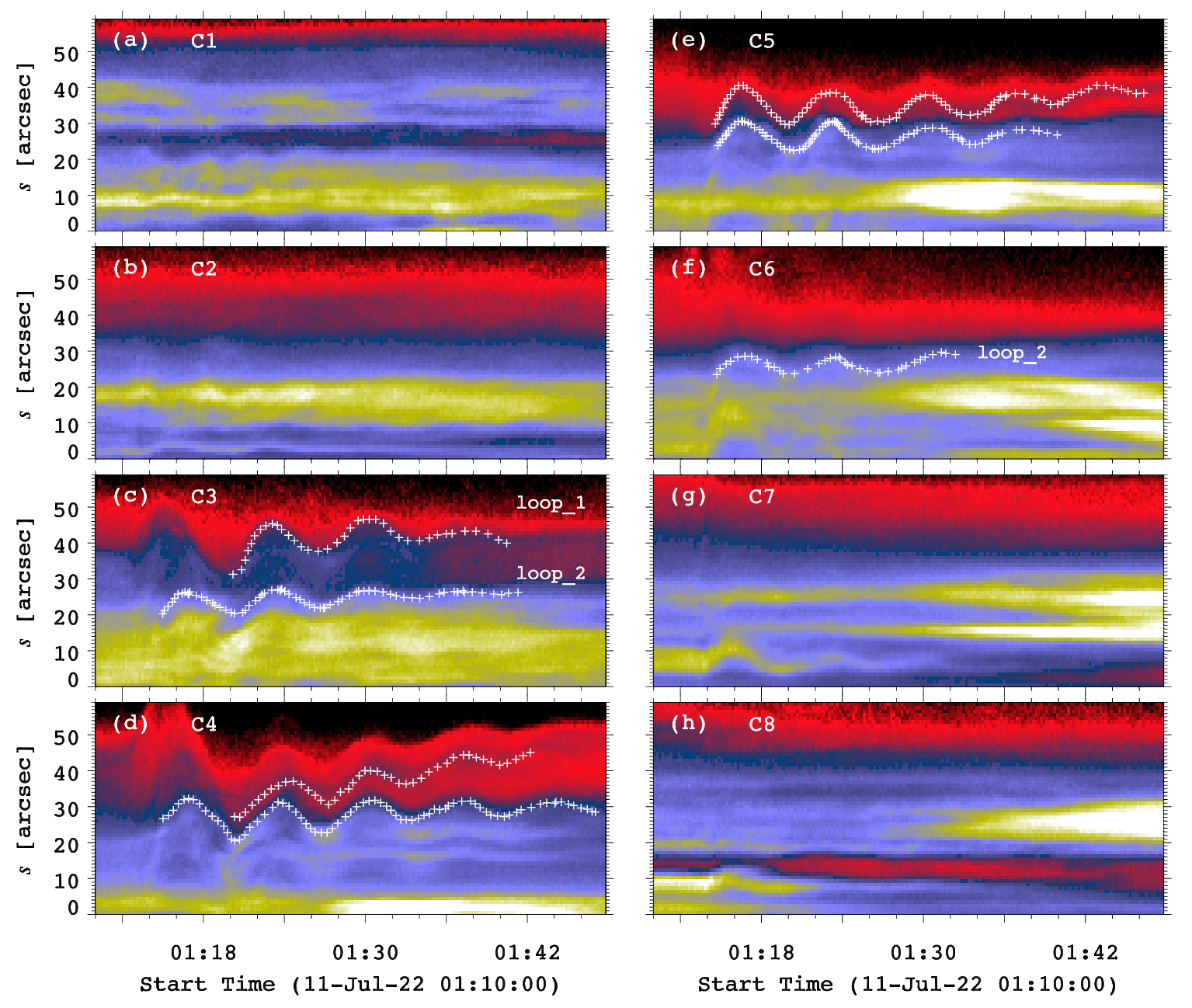}}
\caption{Time-distance diagrams for C1$-$C8 in 171\,{\AA}.
The white ``+" symbols represent the manually tracked loop positions during the oscillations.}
\label{fig8}
\end{figure}

The AIA 171 {\AA} difference image at 01:26:57 UT is displayed in Figure~\ref{fig7}. 
We select eight slices (C1$-$C8) with the same length of 60$\arcsec$. C1 is at the right leg, C3 is close to the loop top, and C8 is close to the left footpoint.
Time-distance diagrams of the eight slices are displayed in Figure~\ref{fig8}. 
In each panel, the white ``+" symbols represent the loop positions tracked manually.
Two oscillating loop strands, including the higher one (loop\_1) and the lower one (loop\_2), are distinctly identified in the diagrams of C3$-$C6.
It is obvious that as soon as the jet-related flare occurs from beneath, the loops first expand upward, then shrink and oscillate. 
The transverse oscillations last for 3 $-$ 5 cycles with or without attenuation until 01:48 UT.

\begin{figure} 
\centerline{\includegraphics[width=0.7\textwidth,clip=]{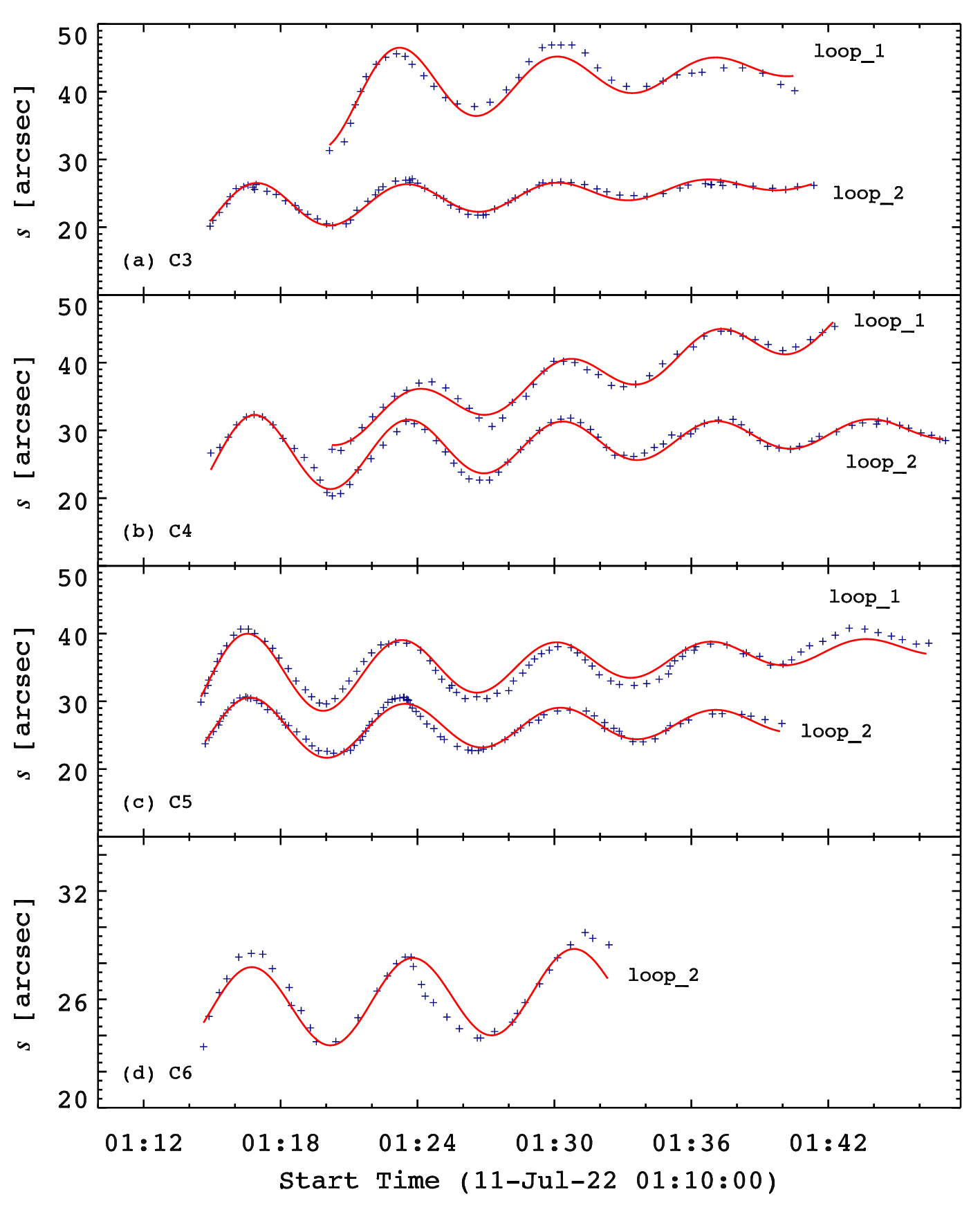}}
\caption{Trajectories of the two loop strands (dark blue ``+" symbols) at C3$-$C6 and the results of curve fittings using Equation~\ref{eqn-1} (\textit{red lines}).}
\label{fig9}
\end{figure}

In Figure~\ref{fig9}, the trajectories of the loop strands are plotted with dark blue ``+" symbols.
To obtain the physical parameters of the kink oscillations, the trajectories are fitted with a damping sine function \citep{nis13}:
\begin{equation} \label{eqn-1}
  y(t)=A_{0}\sin\left[\frac{2\pi}{P}(t-t_{0})+\phi_{0}\right] \mathrm{e}^{-(t-t_{0})/\tau}+y_{0}+k(t-t_{0}),
\end{equation}
where $A_{0}$, $\phi_{0}$, and $y_{0}$ represent the initial amplitude, phase, and displacement at $t_0$.
$P$ and $\tau$ represent the period and damping time of kink oscillation. $k$ denotes the linear drift speed of coronal loops.
The curve fittings are performed using the standard routine \textsf{mpfit.pro} in SSW, and the parameters are listed in Table~\ref{tab-3}.
Since C3 is close to the loop tops, the two loop strands are $\approx$ 11\,Mm apart before the oscillations. For the higher strand (loop\_1), 
the initial amplitude reaches up to $\approx$ 5.75\,Mm at C3. 
The period is between $\approx$ 396\,s and $\approx$ 413\,s, with an average value of $\approx$ 405\,s, 
which is close to the period of the transverse loop oscillation excited by a non-radial flux rope eruption on 7 December 2012 \citep{zqm22a}.
\citet{god16a} carried out a statistical study of 58 decaying kink oscillations observed by SDO/AIA between 2010 and 2014.
The initial loop displacements and oscillation amplitudes are analyzed in detail.
Although the initial loop displacement prescribes the initial amplitude of the oscillation in general, 
there are cases when the initial loop displacement exceeds the initial amplitude of the oscillation.
In the current study, the initial loop displacements at C3 and C4 during 01:13$-$01:20 UT are too large to perform a coherent curve fitting between 01:13 UT and 01:42 UT (see Figure~\ref{fig8}c-d).
Accordingly, we fit the oscillations of upper strands of C3 and C4 after 01:20 UT, which result in lower values of $A_{0}$.
For the lower strand (loop\_2), the initial amplitude reaches $\approx$ 4.9 Mm at C4. 
The period is between $\approx$ 394\,s and $\approx$ 424\,s, with an average value of $\approx$ 407\,s.
For each strand, the initial phase has marginal variation across the strand, suggesting that the whole strand oscillates in phase (see fourth column of Table~\ref{tab-3}).
Besides, the amplitudes are maximal at the loop tops and are negligible close to the footpoints. Therefore, the kink oscillations of the loop strands are of fundamental standing mode.
Both strands show attenuation with the damping ratio $\frac{\tau}{P}$ between $\approx$ 1.7 and $\approx$ 3.2 in most cases.
The oscillations are considered to be decayless for very small amplitudes (loop\_1 at C4 and loop\_2 at C6).
The value of $k$ is between 0.7 and 8.1\,km s$^{-1}$, meaning slow expansions during the oscillations \citep{zqm22b,zqm23}.
Timeline of the whole events is displayed in Table~\ref{tab-4}.

\begin{table}
\caption{Physical parameters of kink oscillations of the two loop strands (loop\_1 and loop\_2).
$A_{0}$, $\phi_{0}$, and $y_{0}$ denote the initial amplitude, phase, and displacement at $t_{0}$.
$P$ and $\tau$ represent the period and damping time. $k$ denotes drift speed of the loops during the oscillations.}
\label{tab-3}
\tabcolsep 1.5mm
\begin{tabular}{cccccccccc}
\hline
Slice& $t_0$ & $A_0$ & $\phi_0$ & $P$ & $\tau$ & $\frac{\tau}{P}$ & Type & $y_0$ & $k$ \\  
     & [UT] & [Mm] & [rad] & [s] & [s] & - & - & [Mm] & [km\,s$^{-1}$] \\
\hline
C3\_loop\_1& 01:20:08 & 5.75 & 4.99 & 412.8 & 721.6 & 1.75 & decaying & 28.8 & 2.4 \\
C4\_loop\_1& 01:20:15 & 2.15 & 4.39 & 395.6 &47733.6 &120.66 & decayless & 22.2 & 8.1 \\
C5\_loop\_1& 01:14:31 & 5.25 & 5.93 & 405.4 & 1116.3 & 2.75 & decaying & 24.1 & 1.9 \\
\hline
C3\_loop\_2& 01:14:55 & 3.04 & 5.88 & 394.2 & 1007.0 & 2.55 & decaying & 16.3 & 1.9 \\
C4\_loop\_2& 01:14:56 & 4.86 & 6.03 & 404.0 & 1309.8 & 3.24 & decaying & 18.8 & 1.7 \\
C5\_loop\_2& 01:14:42 & 3.90 & 6.01 & 407.9 & 1293.7 & 3.17 & decaying & 18.5 & 0.7 \\
C6\_loop\_2& 01:14:38 & 1.67 & 6.01 & 423.8 & 38302.1 & 90.38 & decayless & 18.4 & 0.9 \\
\hline
\end{tabular}
\end{table}

\begin{table}
\caption{Timeline of the whole events.}
\label{tab-4}
\tabcolsep 1.5mm
\begin{tabular}{cc}
\hline
Time (UT) & Activity  \\
\hline
01:08 & Start time of the flare \\
01:10 & Start time of the coronal jet \\
01:12 & Peak time of the flare \\
01:12 & Start time of the loop oscillations \\
01:16 & End time of the flare \\
01:34 & End time of the jet \\
01:48 & End time of the loop oscillations \\
\hline
\end{tabular}
\end{table}

\section{Discussion} \label{discussion}

\subsection{Coronal Seismology}
Coronal seismology is a powerful method to diagnose the magnetic field strength of the oscillating loops, which are difficult to measure directly \citep{van08,yzh20,yzh24}.
The period of the kink oscillation of the standing mode is \citep{naka21}:
\begin{equation} \label{eqn-2}
  P=\frac{2L}{nC_k},
\end{equation}
where $L$ is the loop length and $n$ represents the number of harmonics (for the fundamental mode, $n=1$). $C_k$ is the kink speed:
\begin{equation} \label{eqn-3}
  C_k=C_A\sqrt{\frac{2}{1+\rho_{o}/\rho_{i}}},
\end{equation}
where $C_A$ is the internal Alfv\'{e}n speed of the loop. $\rho_{o}$ and $\rho_{i}$ stand for external and internal plasma densities.

In Table~\ref{tab-3}, the average periods of the oscillation for loop\_1 and loop\_2 are $\approx$ 405\,s and $\approx$ 407\,s, respectively.
Owing to the lower resolution of STA/EUVI compared with SDO/AIA, it is difficult to separate the two strands, which are close to each other.
In Figure~\ref{fig5}, the locations of two strands are outlined with white ``+" symbols in AIA 171\,{\AA} (panel a) and EUVI 195\,{\AA} (panel b) at 01:10 UT on 11 July 2022.
The apparent total length of the loop is $\approx$ 250$\arcsec$, which is a lower limit of the loop length in 3D.
Consequently, the lower limits of $C_k$ are 895$_{-17}^{+21}$\,km\,s$^{-1}$ for loop\_1 and 891$_{-35}^{+29}$\,km\,s$^{-1}$ for loop\_2, respectively.
The corresponding lower limits of $C_A$ are estimated to be 664$_{-13}^{+16}$\,km\,s$^{-1}$ and 661$_{-26}^{+22}$\,km\,s$^{-1}$, 
assuming that $\rho_{o}/\rho_{i} \approx 0.1$ \citep{naka01}.

\begin{figure} 
\centerline{\includegraphics[width=0.9\textwidth,clip=]{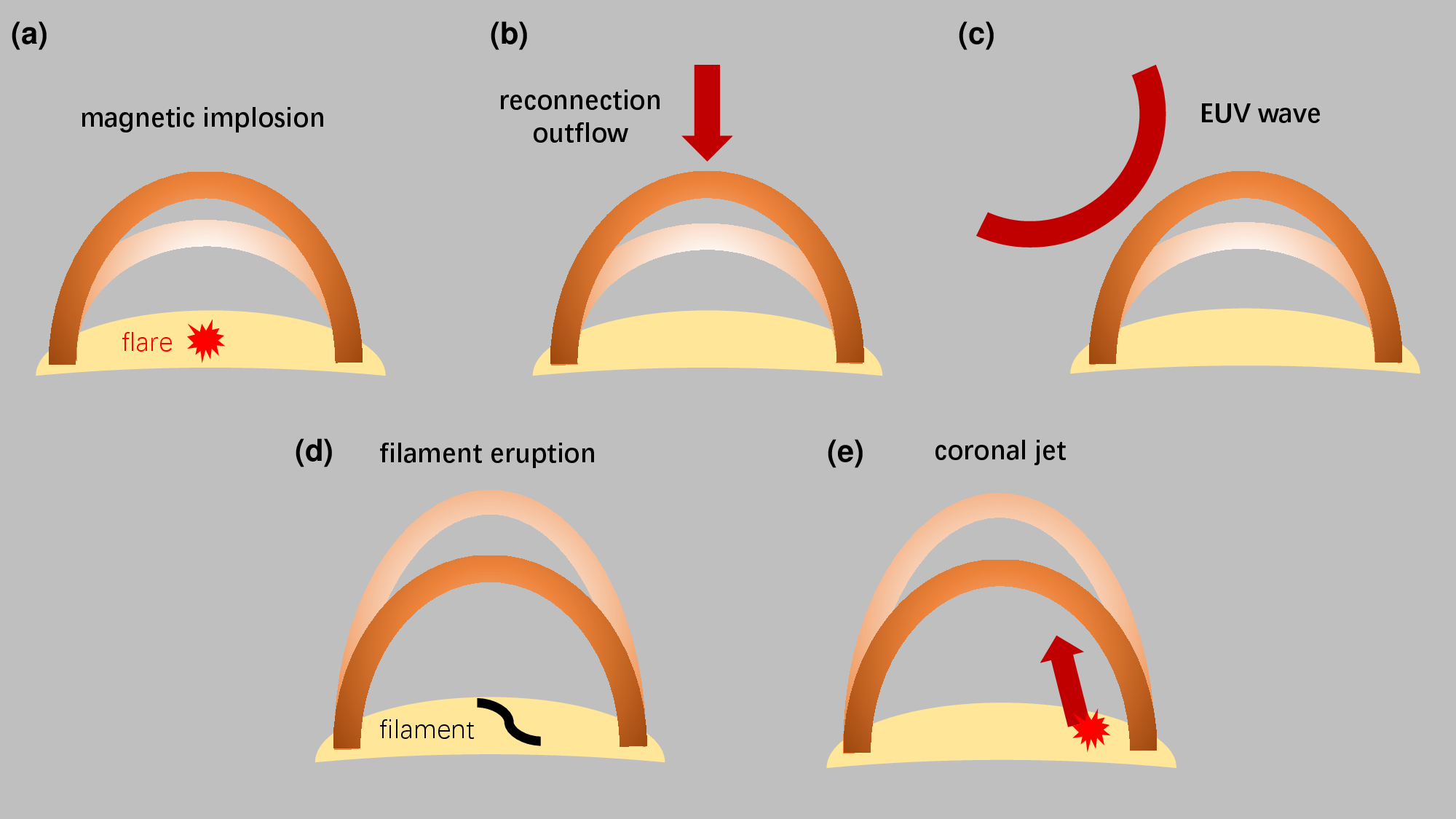}}
\caption{A schematic cartoon to illustrate five triggering mechanisms of kink oscillations of coronal loops:
magnetic implosion (a), reconnection outflows (b), EUV waves (c), filament eruptions (d), and jet-related flares (e).
In each panel, the dark brown loop and light orange loop represent the positions before and after a perturbation, respectively.}
\label{fig10}
\end{figure}

\subsection{Triggering Mechanism of Kink Oscillations}
As mentioned in Section~\ref{intro}, coronal jets are widely spread in the solar atmosphere \citep{liu23}.
A fraction of jets are powerful enough to excite waves and oscillations when interacting with the surrounding magnetic system, such as loop oscillations \citep{sar16,dai21}, 
filament oscillations \citep{luna14,zqm17,ni22,tan23}, EUV waves \citep{shen18,hou23,zqm24},
and quasiperiodic fast-propagating wave trains \citep{zhou24}. \citet{zim15} analyzed 58 kink-oscillation events observed by SDO/AIA, finding that 57 events are accompanied by LCEs. 
In their schematic cartoon, a coronal loop is pushed aside by the LCE. Then, the loop returns back and oscillates in the horizontal direction without an expansion or a contraction.

Using multi-instrument observations and 3D reconstruction of coronal loops \citep{ver10}, \citet{nis17} explored the MHD waves in a loop bundle induced by a failed or confined flare eruption on 24 January 2015. 
The kink oscillations with strong attenuation are vertically polarized with a period of 3.5 $-$ 4\,minutes and an initial amplitude of $\approx$ 5\,Mm.
Since we are unable to determine the polarization of the transverse oscillations, we assume that the oscillations are mainly in the vertical direction.
In Figure~\ref{fig10}, the five panels illustrate main triggering mechanisms of kink oscillations of coronal loops.
In panel a, a flare occurs without a jet beneath the large-scale, overlying loops in the same AR. 
Magnetic implosion, i.e., an impulsive decrease of magnetic pressure due to a rapid release of free energy, 
results in an imbalance of forces acting on the loops and a downward motion as well as an oscillation \citep{sim13,rus15,dud16}.
In panel b, hot reconnection outflow ejects from a flare current sheet, propagates downward, 
and collides with the post-flare loops, generating simultaneous shrinkage and oscillation \citep{wht12b,tian14a,lid17,ree20}.
In panel c, an EUV wave arrives at and pushes down a low-lying coronal loop impulsively, causing a vertical oscillation during the gradual expansion (see Fig. 12 in \citet{zqm22b} for details).
Therefore, the direction of the initial perturbations is downward in the above three cases.
It is noticed that cool and dense condensations (e.g., coronal rains) in coronal loops due to the thermal instability is capable of exciting small-amplitude, vertical loop oscillations \citep{koh17}.
The overall trend of the loop movement is a fast contraction followed by a slow expansion, which is similar to the case of kink oscillations excited by EUV waves \citep{zqm23}.
In panel d, a filament lifts off from below and pushes up a large-scale coronal loop, causing an expansion and oscillation \citep{mro11}.
Meanwhile, the overlying loop prevents the filament from a successful eruption to generate a CME, so that the filament hangs up or returns back to the solar surface.
It is noted that there is no need to interact with the loop closely.
In the event on 14 July 2004, the initial height of the loop is $\approx$ 85\,Mm, while the rising filament stops at a height of $\approx$ 52\,Mm before returning \citep{mro11}.
There is a gap of $\approx$ 33\,Mm between the top of filament and the oscillating loop.
In panel e, beneath the coronal loop, a flare occurs, which is associated with a blowout coronal jet.
The jet returns back after reaching its apex and the falling process is more evident in 304 {\AA} (see Figure~\ref{fig6}).
Transverse oscillations of the loop are excited by the flare \citep{asch99,naka99}.
Accordingly, the direction of the initial perturbations is upward in the above two cases (panels d and e).
The main difference between the cases in panel a and panel e lies in the direction of initial perturbations.
In panel a, the loop moves downward and oscillates as a result of the magnetic implosion generated by the flare.
The oscillation is accompanied by a slow contraction \citep[][see their Fig. 4c]{rus15}.
In contrast, the loop moves upward and oscillates as a result of a jet-related, confined flare \citep{dai21}.
The oscillation is accompanied by a slow expansion (see Figure~\ref{fig9} and Table~\ref{tab-3}).
From this point of view, the flare-related jet may play an important role in determining the direction of the initial perturbation of the oscillating loop.

\begin{figure} 
\centerline{\includegraphics[width=0.9\textwidth,clip=]{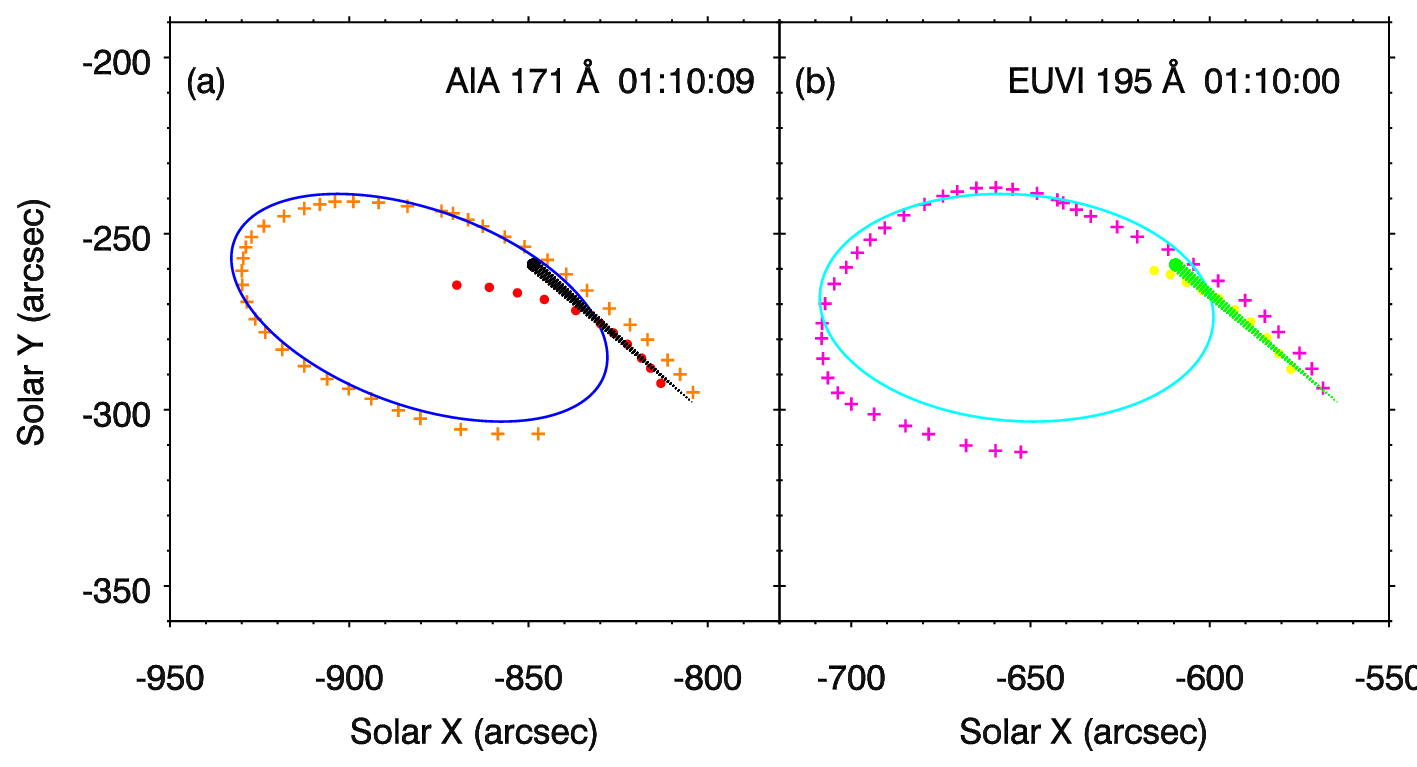}}
\caption{The oscillating loop (orange and magenta plus symbols) and jet spire (red and yellow dots) observed by AIA (left panel) and EUVI (right panel), respectively.
Projections of the reconstructed loop (blue and cyan ovals) and jet spire (black and green dots) are superposed.}
\label{fig11}
\end{figure}

In Figure~\ref{fig11}, the oscillating loop observed by SDO/AIA 171 {\AA} and STA/EUVI 195 {\AA} at 01:10 UT are drawn with orange and magenta plus symbols 
in the left and right panels (see also Figure~\ref{fig5}a-b). 
To derive the 3D geometry of the loop, we use an ellipse or a circle to fit the loop simultaneously observed from two vantage points.
Projections of the reconstructed 3D loop (a circle with a radius of 55$\arcsec$) on AIA and EUVI FOVs are drawn with blue and cyan ovals.
It is seen that the fitting is acceptable for most of the points, although the footpoints of the reconstructed loop is hard to determine.
Likewise, the jet spire observed by AIA 304 {\AA} and EUVI 195 {\AA} are represented by red and yellow dots, respectively.
To derive the real direction of the jet, we apply the revised cone model, which is proposed to investigate non-radial filament eruptions \citep{zhang21,zhang22}.
The tip of the cone is placed at the source location of the eruption, while the direction of eruption is determined by two inclination angles ($\theta_1$, $\phi_1$) 
with respect to the local vertical.
For the coronal jet originating from a minifilament eruption, the angular width of the cone is set to be 6$^{\circ}$.
Projections of the reconstructed 3D jet spire on AIA and EUVI FOVs are drawn with black and green dots, which are in line with the observed jet.
It is obvious that the reconstruction of the jet spire is satisfactory using the revised cone model.

\begin{figure} 
\centerline{\includegraphics[width=0.9\textwidth,clip=]{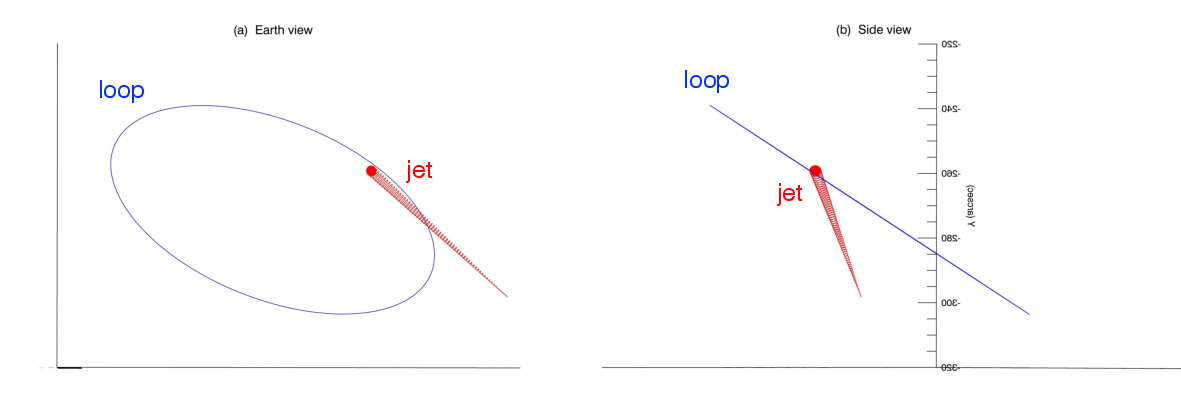}}
\caption{Earth view and side view of the reconstructed loop (blue oval) and jet spire (red dots).}
\label{fig12}
\end{figure}

To investigate whether the jet and loop are coplanar, we showcase the reconstructed loop and jet from two perspectives in Figure~\ref{fig12}.
The left panel shows the loop (blue line) and jet (red dots) from Earth's view, while the right panel shows them from a side view.
It is found that the jet and loop plane has an acute angle of 37$^{\circ}$ in 3D. 
In other words, the propagation of jet still has a significant component within the loop plane to trigger the transverse oscillations as described in Section~\ref{results}.

\section{Summary} \label{summary}
In this article, we report multiwavelength and multiview observations of the transverse oscillations of two loop strands 
induced by a jet-related, confined flare in AR 13056 on 11 July 2022. The main results are summarized as follows:
\begin{enumerate}[(i)]
\item The jet originates close to the right footpoint of the loops and propagates in the northeast direction.
The average rise time and fall time of the jet are $\approx$ 11 and $\approx$ 13.5\,minutes, so that the lifetime of the jet reaches $\approx$ 24.5\,minutes.
The rising motion of the jet is divided into two phases with average velocities of $\approx$ 164 and $\approx$ 546\,km\,s$^{-1}$.
The falling motion of the jet is coherent with an average velocity of $\approx$ 124\,km\,s$^{-1}$.
\item The transverse oscillations of the loops, lasting for 3 $-$ 5 cycles, are of the fundamental standing kink mode. 
Meanwhile,  the oscillations are accompanied by slow expansions.
The maximal initial amplitudes are $\approx$ 5.8 and $\approx$ 4.9\,Mm. The average periods are $\approx$ 405\,s and $\approx$ 407\,s.
The lower limits of the kink speed are 895$_{-17}^{+21}$\,km\,s$^{-1}$ for loop\_1 and 891$_{-35}^{+29}$\,km\,s$^{-1}$ for loop\_2, respectively.
The corresponding lower limits of the Alfv\'{e}n speed are estimated to be 664$_{-13}^{+16}$\,km\,s$^{-1}$ and 661$_{-26}^{+22}$\,km\,s$^{-1}$.
\end{enumerate}

\begin{acks}
The authors appreciate the referee for his/her valuable suggestions to improve the quality of this article.
SDO is a mission of NASA\rq{}s Living With a Star Program. AIA and HMI data are courtesy of the NASA/SDO science teams.
STEREO/SECCHI data are provided by a consortium of US, UK, Germany, Belgium, and France.
\end{acks}

\begin{fundinginformation}
This work is supported by the National Key R\&D Program of China 2021YFA1600500 (2021YFA1600502), 2019YFA0405000, 
the Strategic Priority Research Program of the Chinese Academy of Sciences, Grant No. XDB0560000, 
NSFC under the grant numbers 12373065, 12373057, and 12003072, 
the Natural Science Foundation of Jiangsu Province (BK20231510), the Beijing Natural Science Foundation (Grant No.1222029), 
and the Project Supported by the Specialized Research Fund for State Key Laboratory of Solar Activity and Space Weather, Chinese Academy of Sciences.
\end{fundinginformation}

\bibliographystyle{spr-mp-sola}
\bibliography{ref}

\begin{thebibliography}{120}
\ifx\bisbn     \undefined \def\bisbn  #1{ISBN #1}\fi
\ifx\binits    \undefined \def\binits#1{#1}\fi
\ifx\bauthor   \undefined \def\bauthor#1{#1}\fi
\ifx\batitle   \undefined \def\batitle#1{#1}\fi
\ifx\bjtitle   \undefined \def\bjtitle#1{\textit{#1}}\fi
\ifx\bvolume   \undefined \def\bvolume#1{\textbf{#1}}\fi
\ifx\byear     \undefined \def\byear#1{#1}\fi
\ifx\bissue    \undefined \def\bissue#1{#1}\fi
\ifx\bfpage    \undefined \def\bfpage#1{#1}\fi
\ifx\blpage    \undefined \def\blpage #1{#1}\fi
\ifx\burl      \undefined \def\burl#1{#1}\fi
\ifx\href      \undefined \def\href#1#2{#2}\fi
\ifx\betal     \undefined \def\betal{et al.}\fi
\ifx\bctitle   \undefined \def\bctitle#1{#1}\fi
\ifx\beditor   \undefined \def\beditor#1{#1}\fi
\ifx\bbtitle   \undefined \def\bbtitle#1{\textit{#1}}\fi
\ifx\bedition  \undefined \def\bedition#1{#1}\fi
\ifx\bseriesno \undefined \def\bseriesno#1{\textbf{#1}}\fi
\ifx\blocation \undefined \def\blocation#1{#1}\fi
\ifx\bsertitle \undefined \def\bsertitle#1{\textit{#1}}\fi
\ifx\bsnm      \undefined \def\bsnm#1{#1}\fi
\ifx\bsuffix   \undefined \def\bsuffix#1{#1}\fi
\ifx\bparticle \undefined \def\bparticle#1{#1}\fi
\ifx\barticle  \undefined \def\barticle#1{}\fi
\ifx\binstitute  \undefined \def\binstitute#1{#1}\fi
\ifx\bpublisher  \undefined \def\bpublisher#1{#1}\fi
\ifx\doiurl    \undefined \def\doiurl#1{\href{#1}{DOI}}\fi
\makeatletter
\def\safeHref#1#2#3{\in@{http}{#2}\ifin@\href{#2}{#3}\else\href{#1#2}{#3}\fi}
\makeatother
\ifx\adsurl    \undefined
  \def\adsurl#1{\safeHref{https://ui.adsabs.harvard.edu/abs/}{#1}{ADS}}\fi
\ifx\arxivurl  \undefined
  \def\arxivurl#1{\safeHref{http://arxiv.org/abs/}{#1}{arXiv}}\fi
\ifx\botherref \undefined \def\botherref#1{}\fi
\ifx\url       \undefined \def\url#1{#1}\fi
\ifx\bchapter  \undefined \def\bchapter#1{}\fi
\ifx\bbook     \undefined \def\bbook#1{}\fi
\ifx\bcomment  \undefined \def\bcomment#1{#1}\fi
\ifx\oauthor   \undefined \def\oauthor#1{#1}\fi
\ifx\citeauthoryear \undefined\def \citeauthoryear#1{#1}\fi
\def\endbibitem {}
\ifx\bconflocation  \undefined \def\bconflocation#1{#1} \fi

\bibitem[\protect\citeauthoryear{{Anfinogentov}, {Nakariakov}, and
  {Nistic{\`o}}}{2015}]{anf15}
\begin{barticle}
\bauthor{\bsnm{{Anfinogentov}}, \binits{S.A.}},
\bauthor{\bsnm{{Nakariakov}}, \binits{V.M.}},
\bauthor{\bsnm{{Nistic{\`o}}}, \binits{G.}}:
\byear{2015},
\batitle{{Decayless low-amplitude kink oscillations: a common phenomenon in the
  solar corona?}}
\bjtitle{\aap}
\bvolume{583},
\bfpage{A136}.
\doiurl{https://doi.org/10.1051/0004-6361/201526195}.
\adsurl{2015A&A...583A.136A}.
\end{barticle}
\endbibitem

\bibitem[\protect\citeauthoryear{{Aschwanden} and {Schrijver}}{2011}]{asch11}
\begin{barticle}
\bauthor{\bsnm{{Aschwanden}}, \binits{M.J.}},
\bauthor{\bsnm{{Schrijver}}, \binits{C.J.}}:
\byear{2011},
\batitle{{Coronal Loop Oscillations Observed with Atmospheric Imaging
  Assembly{\textemdash}Kink Mode with Cross-sectional and Density
  Oscillations}}.
\bjtitle{\apj}
\bvolume{736},
\bfpage{102}.
\doiurl{https://doi.org/10.1088/0004-637X/736/2/102}.
\adsurl{2011ApJ...736..102A}.
\end{barticle}
\endbibitem

\bibitem[\protect\citeauthoryear{{Aschwanden} et~al.}{1999}]{asch99}
\begin{barticle}
\bauthor{\bsnm{{Aschwanden}}, \binits{M.J.}},
\bauthor{\bsnm{{Fletcher}}, \binits{L.}},
\bauthor{\bsnm{{Schrijver}}, \binits{C.J.}},
\bauthor{\bsnm{{Alexander}}, \binits{D.}}:
\byear{1999},
\batitle{{Coronal Loop Oscillations Observed with the Transition Region and
  Coronal Explorer}}.
\bjtitle{\apj}
\bvolume{520},
\bfpage{880}.
\doiurl{https://doi.org/10.1086/307502}.
\adsurl{1999ApJ...520..880A}.
\end{barticle}
\endbibitem

\bibitem[\protect\citeauthoryear{{Aschwanden} et~al.}{2002}]{asch02}
\begin{barticle}
\bauthor{\bsnm{{Aschwanden}}, \binits{M.J.}},
\bauthor{\bsnm{{De Pontieu}}, \binits{B.}},
\bauthor{\bsnm{{Schrijver}}, \binits{C.J.}},
\bauthor{\bsnm{{Title}}, \binits{A.M.}}:
\byear{2002},
\batitle{{Transverse Oscillations in Coronal Loops Observed with TRACE II.
  Measurements of Geometric and Physical Parameters}}.
\bjtitle{\solphys}
\bvolume{206},
\bfpage{99}.
\doiurl{https://doi.org/10.1023/A:1014916701283}.
\adsurl{2002SoPh..206...99A}.
\end{barticle}
\endbibitem

\bibitem[\protect\citeauthoryear{{Beckers}}{1972}]{bec72}
\begin{barticle}
\bauthor{\bsnm{{Beckers}}, \binits{J.M.}}:
\byear{1972},
\batitle{{Solar Spicules}}.
\bjtitle{\araa}
\bvolume{10},
\bfpage{73}.
\doiurl{https://doi.org/10.1146/annurev.aa.10.090172.000445}.
\adsurl{1972ARA&A..10...73B}.
\end{barticle}
\endbibitem

\bibitem[\protect\citeauthoryear{{Chae} et~al.}{1999}]{chae99}
\begin{barticle}
\bauthor{\bsnm{{Chae}}, \binits{J.}},
\bauthor{\bsnm{{Qiu}}, \binits{J.}},
\bauthor{\bsnm{{Wang}}, \binits{H.}},
\bauthor{\bsnm{{Goode}}, \binits{P.R.}}:
\byear{1999},
\batitle{{Extreme-Ultraviolet Jets and H{\ensuremath{\alpha}} Surges in Solar
  Microflares}}.
\bjtitle{\apjl}
\bvolume{513},
\bfpage{L75}.
\doiurl{https://doi.org/10.1086/311910}.
\adsurl{1999ApJ...513L..75C}.
\end{barticle}
\endbibitem

\bibitem[\protect\citeauthoryear{{Chen}, {Zhang}, and {Ma}}{2012}]{chen12}
\begin{barticle}
\bauthor{\bsnm{{Chen}}, \binits{H.-D.}},
\bauthor{\bsnm{{Zhang}}, \binits{J.}},
\bauthor{\bsnm{{Ma}}, \binits{S.-L.}}:
\byear{2012},
\batitle{{The kinematics of an untwisting solar jet in a polar coronal hole
  observed by SDO/AIA}}.
\bjtitle{Research in Astronomy and Astrophysics}
\bvolume{12},
\bfpage{573}.
\doiurl{https://doi.org/10.1088/1674-4527/12/5/009}.
\adsurl{2012RAA....12..573C}.
\end{barticle}
\endbibitem

\bibitem[\protect\citeauthoryear{{Chen} et~al.}{2015}]{chen15}
\begin{barticle}
\bauthor{\bsnm{{Chen}}, \binits{J.}},
\bauthor{\bsnm{{Su}}, \binits{J.}},
\bauthor{\bsnm{{Yin}}, \binits{Z.}},
\bauthor{\bsnm{{Priya}}, \binits{T.G.}},
\bauthor{\bsnm{{Zhang}}, \binits{H.}},
\bauthor{\bsnm{{Liu}}, \binits{J.}},
\bauthor{\bsnm{{Xu}}, \binits{H.}},
\bauthor{\bsnm{{Yu}}, \binits{S.}}:
\byear{2015},
\batitle{{Recurrent Solar Jets Induced by a Satellite Spot and Moving Magnetic
  Features}}.
\bjtitle{\apj}
\bvolume{815},
\bfpage{71}.
\doiurl{https://doi.org/10.1088/0004-637X/815/1/71}.
\adsurl{2015ApJ...815...71C}.
\end{barticle}
\endbibitem

\bibitem[\protect\citeauthoryear{{Chen} et~al.}{2017}]{chen17}
\begin{barticle}
\bauthor{\bsnm{{Chen}}, \binits{J.}},
\bauthor{\bsnm{{Su}}, \binits{J.}},
\bauthor{\bsnm{{Deng}}, \binits{Y.}},
\bauthor{\bsnm{{Priest}}, \binits{E.R.}}:
\byear{2017},
\batitle{{A Complex Solar Coronal Jet with Two Phases}}.
\bjtitle{\apj}
\bvolume{840},
\bfpage{54}.
\doiurl{https://doi.org/10.3847/1538-4357/aa6c59}.
\adsurl{2017ApJ...840...54C}.
\end{barticle}
\endbibitem

\bibitem[\protect\citeauthoryear{{Chen} et~al.}{2022}]{chen22}
\begin{barticle}
\bauthor{\bsnm{{Chen}}, \binits{J.}},
\bauthor{\bsnm{{Erd{\'e}lyi}}, \binits{R.}},
\bauthor{\bsnm{{Liu}}, \binits{J.}},
\bauthor{\bsnm{{Deng}}, \binits{Y.}},
\bauthor{\bsnm{{Dover}}, \binits{F.M.}},
\bauthor{\bsnm{{Zhang}}, \binits{Q.}},
\bauthor{\bsnm{{Zhang}}, \binits{M.}},
\bauthor{\bsnm{{Li}}, \binits{L.}},
\bauthor{\bsnm{{Su}}, \binits{J.}}:
\byear{2022},
\batitle{{Blobs in a Solar EUV Jet}}.
\bjtitle{Frontiers in Astronomy and Space Sciences}
\bvolume{8},
\bfpage{238}.
\doiurl{https://doi.org/10.3389/fspas.2021.786856}.
\adsurl{2022FrASS...8..238C}.
\end{barticle}
\endbibitem

\bibitem[\protect\citeauthoryear{{Cirtain} et~al.}{2007}]{cir07}
\begin{barticle}
\bauthor{\bsnm{{Cirtain}}, \binits{J.W.}},
\bauthor{\bsnm{{Golub}}, \binits{L.}},
\bauthor{\bsnm{{Lundquist}}, \binits{L.}},
\bauthor{\bsnm{{van Ballegooijen}}, \binits{A.}},
\bauthor{\bsnm{{Savcheva}}, \binits{A.}},
\bauthor{\bsnm{{Shimojo}}, \binits{M.}},
\bauthor{\bsnm{{DeLuca}}, \binits{E.}},
\bauthor{\bsnm{{Tsuneta}}, \binits{S.}},
\bauthor{\bsnm{{Sakao}}, \binits{T.}},
\bauthor{\bsnm{{Reeves}}, \binits{K.}},
\bauthor{\bsnm{{Weber}}, \binits{M.}},
\bauthor{\bsnm{{Kano}}, \binits{R.}},
\bauthor{\bsnm{{Narukage}}, \binits{N.}},
\bauthor{\bsnm{{Shibasaki}}, \binits{K.}}:
\byear{2007},
\batitle{{Evidence for Alfv{\'e}n Waves in Solar X-ray Jets}}.
\bjtitle{Science}
\bvolume{318},
\bfpage{1580}.
\doiurl{https://doi.org/10.1126/science.1147050}.
\adsurl{2007Sci...318.1580C}.
\end{barticle}
\endbibitem

\bibitem[\protect\citeauthoryear{{Dai} et~al.}{2021}]{dai21}
\begin{barticle}
\bauthor{\bsnm{{Dai}}, \binits{J.}},
\bauthor{\bsnm{{Zhang}}, \binits{Q.M.}},
\bauthor{\bsnm{{Su}}, \binits{Y.N.}},
\bauthor{\bsnm{{Ji}}, \binits{H.S.}}:
\byear{2021},
\batitle{{Transverse oscillation of a coronal loop induced by a flare-related
  jet}}.
\bjtitle{\aap}
\bvolume{646},
\bfpage{A12}.
\doiurl{https://doi.org/10.1051/0004-6361/202039013}.
\adsurl{2021A&A...646A..12D}.
\end{barticle}
\endbibitem

\bibitem[\protect\citeauthoryear{{De Pontieu} et~al.}{2007}]{dp07}
\begin{barticle}
\bauthor{\bsnm{{De Pontieu}}, \binits{B.}},
\bauthor{\bsnm{{McIntosh}}, \binits{S.}},
\bauthor{\bsnm{{Hansteen}}, \binits{V.H.}},
\bauthor{\bsnm{{Carlsson}}, \binits{M.}},
\bauthor{\bsnm{{Schrijver}}, \binits{C.J.}},
\bauthor{\bsnm{{Tarbell}}, \binits{T.D.}},
\bauthor{\bsnm{{Title}}, \binits{A.M.}},
\bauthor{\bsnm{{Shine}}, \binits{R.A.}},
\bauthor{\bsnm{{Suematsu}}, \binits{Y.}},
\bauthor{\bsnm{{Tsuneta}}, \binits{S.}},
\bauthor{\bsnm{{Katsukawa}}, \binits{Y.}},
\bauthor{\bsnm{{Ichimoto}}, \binits{K.}},
\bauthor{\bsnm{{Shimizu}}, \binits{T.}},
\bauthor{\bsnm{{Nagata}}, \binits{S.}}:
\byear{2007},
\batitle{{A Tale of Two Spicules: The Impact of Spicules on the Magnetic
  Chromosphere}}.
\bjtitle{\pasj}
\bvolume{59},
\bfpage{S655}.
\doiurl{https://doi.org/10.1093/pasj/59.sp3.S655}.
\adsurl{2007PASJ...59S.655D}.
\end{barticle}
\endbibitem

\bibitem[\protect\citeauthoryear{{Duan} et~al.}{2024}]{duan24}
\begin{barticle}
\bauthor{\bsnm{{Duan}}, \binits{Y.}},
\bauthor{\bsnm{{Tian}}, \binits{H.}},
\bauthor{\bsnm{{Chen}}, \binits{H.}},
\bauthor{\bsnm{{Shen}}, \binits{Y.}},
\bauthor{\bsnm{{Sun}}, \binits{Z.}},
\bauthor{\bsnm{{Hou}}, \binits{Z.}},
\bauthor{\bsnm{{Li}}, \binits{C.}}:
\byear{2024},
\batitle{{Formation of Fan-spine Magnetic Topology through Flux Emergence and
  Subsequent Jet Production}}.
\bjtitle{\apjl}
\bvolume{962},
\bfpage{L38}.
\doiurl{https://doi.org/10.3847/2041-8213/ad24f3}.
\adsurl{2024ApJ...962L..38D}.
\end{barticle}
\endbibitem

\bibitem[\protect\citeauthoryear{{Dud{\'\i}k} et~al.}{2016}]{dud16}
\begin{barticle}
\bauthor{\bsnm{{Dud{\'\i}k}}, \binits{J.}},
\bauthor{\bsnm{{Polito}}, \binits{V.}},
\bauthor{\bsnm{{Janvier}}, \binits{M.}},
\bauthor{\bsnm{{Mulay}}, \binits{S.M.}},
\bauthor{\bsnm{{Karlick{\'y}}}, \binits{M.}},
\bauthor{\bsnm{{Aulanier}}, \binits{G.}},
\bauthor{\bsnm{{Del Zanna}}, \binits{G.}},
\bauthor{\bsnm{{Dzif{\v{c}}{\'a}kov{\'a}}}, \binits{E.}},
\bauthor{\bsnm{{Mason}}, \binits{H.E.}},
\bauthor{\bsnm{{Schmieder}}, \binits{B.}}:
\byear{2016},
\batitle{{Slipping Magnetic Reconnection, Chromospheric Evaporation, Implosion,
  and Precursors in the 2014 September 10 X1.6-Class Solar Flare}}.
\bjtitle{\apj}
\bvolume{823},
\bfpage{41}.
\doiurl{https://doi.org/10.3847/0004-637X/823/1/41}.
\adsurl{2016ApJ...823...41D}.
\end{barticle}
\endbibitem

\bibitem[\protect\citeauthoryear{{Gao} et~al.}{2022}]{gao22}
\begin{barticle}
\bauthor{\bsnm{{Gao}}, \binits{Y.}},
\bauthor{\bsnm{{Tian}}, \binits{H.}},
\bauthor{\bsnm{{Van Doorsselaere}}, \binits{T.}},
\bauthor{\bsnm{{Chen}}, \binits{Y.}}:
\byear{2022},
\batitle{{Decayless Oscillations in Solar Coronal Bright Points}}.
\bjtitle{\apj}
\bvolume{930},
\bfpage{55}.
\doiurl{https://doi.org/10.3847/1538-4357/ac62cf}.
\adsurl{2022ApJ...930...55G}.
\end{barticle}
\endbibitem

\bibitem[\protect\citeauthoryear{{Garcia}}{1994}]{gar94}
\begin{barticle}
\bauthor{\bsnm{{Garcia}}, \binits{H.A.}}:
\byear{1994},
\batitle{{Temperature and Emission Measure from Goes Soft X-Ray Measurements}}.
\bjtitle{\solphys}
\bvolume{154},
\bfpage{275}.
\doiurl{https://doi.org/10.1007/BF00681100}.
\adsurl{1994SoPh..154..275G}.
\end{barticle}
\endbibitem

\bibitem[\protect\citeauthoryear{{Goddard} and {Nakariakov}}{2016}]{god16b}
\begin{barticle}
\bauthor{\bsnm{{Goddard}}, \binits{C.R.}},
\bauthor{\bsnm{{Nakariakov}}, \binits{V.M.}}:
\byear{2016},
\batitle{{Dependence of kink oscillation damping on the amplitude}}.
\bjtitle{\aap}
\bvolume{590},
\bfpage{L5}.
\doiurl{https://doi.org/10.1051/0004-6361/201628718}.
\adsurl{2016A&A...590L...5G}.
\end{barticle}
\endbibitem

\bibitem[\protect\citeauthoryear{{Goddard} et~al.}{2016}]{god16a}
\begin{barticle}
\bauthor{\bsnm{{Goddard}}, \binits{C.R.}},
\bauthor{\bsnm{{Nistic{\`o}}}, \binits{G.}},
\bauthor{\bsnm{{Nakariakov}}, \binits{V.M.}},
\bauthor{\bsnm{{Zimovets}}, \binits{I.V.}}:
\byear{2016},
\batitle{{A statistical study of decaying kink oscillations detected using
  SDO/AIA}}.
\bjtitle{\aap}
\bvolume{585},
\bfpage{A137}.
\doiurl{https://doi.org/10.1051/0004-6361/201527341}.
\adsurl{2016A&A...585A.137G}.
\end{barticle}
\endbibitem

\bibitem[\protect\citeauthoryear{{Gosain}}{2012}]{gos12}
\begin{barticle}
\bauthor{\bsnm{{Gosain}}, \binits{S.}}:
\byear{2012},
\batitle{{Evidence for Collapsing Fields in the Corona and Photosphere during
  the 2011 February 15 X2.2 Flare: SDO/AIA and HMI Observations}}.
\bjtitle{\apj}
\bvolume{749},
\bfpage{85}.
\doiurl{https://doi.org/10.1088/0004-637X/749/1/85}.
\adsurl{2012ApJ...749...85G}.
\end{barticle}
\endbibitem

\bibitem[\protect\citeauthoryear{{Handy} et~al.}{1999}]{han99}
\begin{barticle}
\bauthor{\bsnm{{Handy}}, \binits{B.N.}},
\bauthor{\bsnm{{Acton}}, \binits{L.W.}},
\bauthor{\bsnm{{Kankelborg}}, \binits{C.C.}},
\bauthor{\bsnm{{Wolfson}}, \binits{C.J.}},
\bauthor{\bsnm{{Akin}}, \binits{D.J.}},
\bauthor{\bsnm{{Bruner}}, \binits{M.E.}},
\bauthor{\bsnm{{Caravalho}}, \binits{R.}},
\bauthor{\bsnm{{Catura}}, \binits{R.C.}},
\bauthor{\bsnm{{Chevalier}}, \binits{R.}},
\bauthor{\bsnm{{Duncan}}, \binits{D.W.}},
\bauthor{\bsnm{{Edwards}}, \binits{C.G.}},
\bauthor{\bsnm{{Feinstein}}, \binits{C.N.}},
\bauthor{\bsnm{{Freeland}}, \binits{S.L.}},
\bauthor{\bsnm{{Friedlaender}}, \binits{F.M.}},
\bauthor{\bsnm{{Hoffmann}}, \binits{C.H.}},
\bauthor{\bsnm{{Hurlburt}}, \binits{N.E.}},
\bauthor{\bsnm{{Jurcevich}}, \binits{B.K.}},
\bauthor{\bsnm{{Katz}}, \binits{N.L.}},
\bauthor{\bsnm{{Kelly}}, \binits{G.A.}},
\bauthor{\bsnm{{Lemen}}, \binits{J.R.}},
\bauthor{\bsnm{{Levay}}, \binits{M.}},
\bauthor{\bsnm{{Lindgren}}, \binits{R.W.}},
\bauthor{\bsnm{{Mathur}}, \binits{D.P.}},
\bauthor{\bsnm{{Meyer}}, \binits{S.B.}},
\bauthor{\bsnm{{Morrison}}, \binits{S.J.}},
\bauthor{\bsnm{{Morrison}}, \binits{M.D.}},
\bauthor{\bsnm{{Nightingale}}, \binits{R.W.}},
\bauthor{\bsnm{{Pope}}, \binits{T.P.}},
\bauthor{\bsnm{{Rehse}}, \binits{R.A.}},
\bauthor{\bsnm{{Schrijver}}, \binits{C.J.}},
\bauthor{\bsnm{{Shine}}, \binits{R.A.}},
\bauthor{\bsnm{{Shing}}, \binits{L.}},
\bauthor{\bsnm{{Strong}}, \binits{K.T.}},
\bauthor{\bsnm{{Tarbell}}, \binits{T.D.}},
\bauthor{\bsnm{{Title}}, \binits{A.M.}},
\bauthor{\bsnm{{Torgerson}}, \binits{D.D.}},
\bauthor{\bsnm{{Golub}}, \binits{L.}},
\bauthor{\bsnm{{Bookbinder}}, \binits{J.A.}},
\bauthor{\bsnm{{Caldwell}}, \binits{D.}},
\bauthor{\bsnm{{Cheimets}}, \binits{P.N.}},
\bauthor{\bsnm{{Davis}}, \binits{W.N.}},
\bauthor{\bsnm{{Deluca}}, \binits{E.E.}},
\bauthor{\bsnm{{McMullen}}, \binits{R.A.}},
\bauthor{\bsnm{{Warren}}, \binits{H.P.}},
\bauthor{\bsnm{{Amato}}, \binits{D.}},
\bauthor{\bsnm{{Fisher}}, \binits{R.}},
\bauthor{\bsnm{{Maldonado}}, \binits{H.}},
\bauthor{\bsnm{{Parkinson}}, \binits{C.}}:
\byear{1999},
\batitle{{The transition region and coronal explorer}}.
\bjtitle{\solphys}
\bvolume{187},
\bfpage{229}.
\doiurl{https://doi.org/10.1023/A:1005166902804}.
\adsurl{1999SoPh..187..229H}.
\end{barticle}
\endbibitem

\bibitem[\protect\citeauthoryear{{Harvey} et~al.}{1996}]{har96}
\begin{barticle}
\bauthor{\bsnm{{Harvey}}, \binits{J.W.}},
\bauthor{\bsnm{{Hill}}, \binits{F.}},
\bauthor{\bsnm{{Hubbard}}, \binits{R.P.}},
\bauthor{\bsnm{{Kennedy}}, \binits{J.R.}},
\bauthor{\bsnm{{Leibacher}}, \binits{J.W.}},
\bauthor{\bsnm{{Pintar}}, \binits{J.A.}},
\bauthor{\bsnm{{Gilman}}, \binits{P.A.}},
\bauthor{\bsnm{{Noyes}}, \binits{R.W.}},
\bauthor{\bsnm{{Title}}, \binits{A.M.}},
\bauthor{\bsnm{{Toomre}}, \binits{J.}},
\bauthor{\bsnm{{Ulrich}}, \binits{R.K.}},
\bauthor{\bsnm{{Bhatnagar}}, \binits{A.}},
\bauthor{\bsnm{{Kennewell}}, \binits{J.A.}},
\bauthor{\bsnm{{Marquette}}, \binits{W.}},
\bauthor{\bsnm{{Patron}}, \binits{J.}},
\bauthor{\bsnm{{Saa}}, \binits{O.}},
\bauthor{\bsnm{{Yasukawa}}, \binits{E.}}:
\byear{1996},
\batitle{{The Global Oscillation Network Group (GONG) Project}}.
\bjtitle{Science}
\bvolume{272},
\bfpage{1284}.
\doiurl{https://doi.org/10.1126/science.272.5266.1284}.
\adsurl{1996Sci...272.1284H}.
\end{barticle}
\endbibitem

\bibitem[\protect\citeauthoryear{{Hong} et~al.}{2016}]{hong16}
\begin{barticle}
\bauthor{\bsnm{{Hong}}, \binits{J.}},
\bauthor{\bsnm{{Jiang}}, \binits{Y.}},
\bauthor{\bsnm{{Yang}}, \binits{J.}},
\bauthor{\bsnm{{Yang}}, \binits{B.}},
\bauthor{\bsnm{{Xu}}, \binits{Z.}},
\bauthor{\bsnm{{Xiang}}, \binits{Y.}}:
\byear{2016},
\batitle{{Mini-filament Eruption as the Initiation of a Jet along Coronal
  Loops}}.
\bjtitle{\apj}
\bvolume{830},
\bfpage{60}.
\doiurl{https://doi.org/10.3847/0004-637X/830/2/60}.
\adsurl{2016ApJ...830...60H}.
\end{barticle}
\endbibitem

\bibitem[\protect\citeauthoryear{{Hou} et~al.}{2023}]{hou23}
\begin{barticle}
\bauthor{\bsnm{{Hou}}, \binits{Z.}},
\bauthor{\bsnm{{Tian}}, \binits{H.}},
\bauthor{\bsnm{{Su}}, \binits{W.}},
\bauthor{\bsnm{{Madjarska}}, \binits{M.S.}},
\bauthor{\bsnm{{Chen}}, \binits{H.}},
\bauthor{\bsnm{{Zheng}}, \binits{R.}},
\bauthor{\bsnm{{Bai}}, \binits{X.}},
\bauthor{\bsnm{{Deng}}, \binits{Y.}}:
\byear{2023},
\batitle{{A Type II Radio Burst Driven by a Blowout Jet on the Sun}}.
\bjtitle{\apj}
\bvolume{953},
\bfpage{171}.
\doiurl{https://doi.org/10.3847/1538-4357/ace31b}.
\adsurl{2023ApJ...953..171H}.
\end{barticle}
\endbibitem

\bibitem[\protect\citeauthoryear{{Hou} et~al.}{2024}]{hou24}
\begin{barticle}
\bauthor{\bsnm{{Hou}}, \binits{Z.}},
\bauthor{\bsnm{{Tian}}, \binits{H.}},
\bauthor{\bsnm{{Madjarska}}, \binits{M.S.}},
\bauthor{\bsnm{{Chen}}, \binits{H.}},
\bauthor{\bsnm{{Samanta}}, \binits{T.}},
\bauthor{\bsnm{{Bai}}, \binits{X.}},
\bauthor{\bsnm{{Li}}, \binits{Z.}},
\bauthor{\bsnm{{Su}}, \binits{Y.}},
\bauthor{\bsnm{{Chen}}, \binits{W.}},
\bauthor{\bsnm{{Deng}}, \binits{Y.}}:
\byear{2024},
\batitle{{Numerous bidirectionally propagating plasma blobs near the
  reconnection site of a solar eruption}}.
\bjtitle{\aap}
\bvolume{687},
\bfpage{A190}.
\doiurl{https://doi.org/10.1051/0004-6361/202449765}.
\adsurl{2024A&A...687A.190H}.
\end{barticle}
\endbibitem

\bibitem[\protect\citeauthoryear{{Howard} et~al.}{2008}]{how08}
\begin{barticle}
\bauthor{\bsnm{{Howard}}, \binits{R.A.}},
\bauthor{\bsnm{{Moses}}, \binits{J.D.}},
\bauthor{\bsnm{{Vourlidas}}, \binits{A.}},
\bauthor{\bsnm{{Newmark}}, \binits{J.S.}},
\bauthor{\bsnm{{Socker}}, \binits{D.G.}},
\bauthor{\bsnm{{Plunkett}}, \binits{S.P.}},
\bauthor{\bsnm{{Korendyke}}, \binits{C.M.}},
\bauthor{\bsnm{{Cook}}, \binits{J.W.}},
\bauthor{\bsnm{{Hurley}}, \binits{A.}},
\bauthor{\bsnm{{Davila}}, \binits{J.M.}},
\bauthor{\bsnm{{Thompson}}, \binits{W.T.}},
\bauthor{\bsnm{{St Cyr}}, \binits{O.C.}},
\bauthor{\bsnm{{Mentzell}}, \binits{E.}},
\bauthor{\bsnm{{Mehalick}}, \binits{K.}},
\bauthor{\bsnm{{Lemen}}, \binits{J.R.}},
\bauthor{\bsnm{{Wuelser}}, \binits{J.P.}},
\bauthor{\bsnm{{Duncan}}, \binits{D.W.}},
\bauthor{\bsnm{{Tarbell}}, \binits{T.D.}},
\bauthor{\bsnm{{Wolfson}}, \binits{C.J.}},
\bauthor{\bsnm{{Moore}}, \binits{A.}},
\bauthor{\bsnm{{Harrison}}, \binits{R.A.}},
\bauthor{\bsnm{{Waltham}}, \binits{N.R.}},
\bauthor{\bsnm{{Lang}}, \binits{J.}},
\bauthor{\bsnm{{Davis}}, \binits{C.J.}},
\bauthor{\bsnm{{Eyles}}, \binits{C.J.}},
\bauthor{\bsnm{{Mapson-Menard}}, \binits{H.}},
\bauthor{\bsnm{{Simnett}}, \binits{G.M.}},
\bauthor{\bsnm{{Halain}}, \binits{J.P.}},
\bauthor{\bsnm{{Defise}}, \binits{J.M.}},
\bauthor{\bsnm{{Mazy}}, \binits{E.}},
\bauthor{\bsnm{{Rochus}}, \binits{P.}},
\bauthor{\bsnm{{Mercier}}, \binits{R.}},
\bauthor{\bsnm{{Ravet}}, \binits{M.F.}},
\bauthor{\bsnm{{Delmotte}}, \binits{F.}},
\bauthor{\bsnm{{Auch\`ere}}, \binits{F.}},
\bauthor{\bsnm{{Delaboudini\`ere}}, \binits{J.P.}},
\bauthor{\bsnm{{Bothmer}}, \binits{V.}},
\bauthor{\bsnm{{Deutsch}}, \binits{W.}},
\bauthor{\bsnm{{Wang}}, \binits{D.}},
\bauthor{\bsnm{{Rich}}, \binits{N.}},
\bauthor{\bsnm{{Cooper}}, \binits{S.}},
\bauthor{\bsnm{{Stephens}}, \binits{V.}},
\bauthor{\bsnm{{Maahs}}, \binits{G.}},
\bauthor{\bsnm{{Baugh}}, \binits{R.}},
\bauthor{\bsnm{{McMullin}}, \binits{D.}},
\bauthor{\bsnm{{Carter}}, \binits{T.}}:
\byear{2008},
\batitle{{Sun Earth Connection Coronal and Heliospheric Investigation
  (SECCHI)}}.
\bjtitle{\ssr}
\bvolume{136},
\bfpage{67}.
\doiurl{https://doi.org/10.1007/s11214-008-9341-4}.
\adsurl{2008SSRv..136...67H}.
\end{barticle}
\endbibitem

\bibitem[\protect\citeauthoryear{{Huang} et~al.}{2020}]{hua20}
\begin{barticle}
\bauthor{\bsnm{{Huang}}, \binits{Z.}},
\bauthor{\bsnm{{Zhang}}, \binits{Q.}},
\bauthor{\bsnm{{Xia}}, \binits{L.}},
\bauthor{\bsnm{{Li}}, \binits{B.}},
\bauthor{\bsnm{{Wu}}, \binits{Z.}},
\bauthor{\bsnm{{Fu}}, \binits{H.}}:
\byear{2020},
\batitle{{Heating at the Remote Footpoints as a Brake on Jet Flows along Loops
  in the Solar Atmosphere}}.
\bjtitle{\apj}
\bvolume{897},
\bfpage{113}.
\doiurl{https://doi.org/10.3847/1538-4357/ab96bd}.
\adsurl{2020ApJ...897..113H}.
\end{barticle}
\endbibitem

\bibitem[\protect\citeauthoryear{{Jiang} et~al.}{2007}]{jia07}
\begin{barticle}
\bauthor{\bsnm{{Jiang}}, \binits{Y.C.}},
\bauthor{\bsnm{{Chen}}, \binits{H.D.}},
\bauthor{\bsnm{{Li}}, \binits{K.J.}},
\bauthor{\bsnm{{Shen}}, \binits{Y.D.}},
\bauthor{\bsnm{{Yang}}, \binits{L.H.}}:
\byear{2007},
\batitle{{The H{\ensuremath{\alpha}} surges and EUV jets from magnetic flux
  emergences and cancellations}}.
\bjtitle{\aap}
\bvolume{469},
\bfpage{331}.
\doiurl{https://doi.org/10.1051/0004-6361:20053954}.
\adsurl{2007A&A...469..331J}.
\end{barticle}
\endbibitem

\bibitem[\protect\citeauthoryear{{Joshi} et~al.}{2018}]{jos18}
\begin{barticle}
\bauthor{\bsnm{{Joshi}}, \binits{N.C.}},
\bauthor{\bsnm{{Nishizuka}}, \binits{N.}},
\bauthor{\bsnm{{Filippov}}, \binits{B.}},
\bauthor{\bsnm{{Magara}}, \binits{T.}},
\bauthor{\bsnm{{Tlatov}}, \binits{A.G.}}:
\byear{2018},
\batitle{{Flux rope breaking and formation of a rotating blowout jet}}.
\bjtitle{\mnras}
\bvolume{476},
\bfpage{1286}.
\doiurl{https://doi.org/10.1093/mnras/sty322}.
\adsurl{2018MNRAS.476.1286J}.
\end{barticle}
\endbibitem

\bibitem[\protect\citeauthoryear{{Kaiser} et~al.}{2008}]{kai08}
\begin{barticle}
\bauthor{\bsnm{{Kaiser}}, \binits{M.L.}},
\bauthor{\bsnm{{Kucera}}, \binits{T.A.}},
\bauthor{\bsnm{{Davila}}, \binits{J.M.}},
\bauthor{\bsnm{{St. Cyr}}, \binits{O.C.}},
\bauthor{\bsnm{{Guhathakurta}}, \binits{M.}},
\bauthor{\bsnm{{Christian}}, \binits{E.}}:
\byear{2008},
\batitle{{The STEREO Mission: An Introduction}}.
\bjtitle{\ssr}
\bvolume{136},
\bfpage{5}.
\doiurl{https://doi.org/10.1007/s11214-007-9277-0}.
\adsurl{2008SSRv..136....5K}.
\end{barticle}
\endbibitem

\bibitem[\protect\citeauthoryear{{Kim}, {Nakariakov}, and {Cho}}{2014}]{kim14}
\begin{barticle}
\bauthor{\bsnm{{Kim}}, \binits{S.}},
\bauthor{\bsnm{{Nakariakov}}, \binits{V.M.}},
\bauthor{\bsnm{{Cho}}, \binits{K.-S.}}:
\byear{2014},
\batitle{{Vertical Kink Oscillation of a Magnetic Flux Rope Structure in the
  Solar Corona}}.
\bjtitle{\apjl}
\bvolume{797},
\bfpage{L22}.
\doiurl{https://doi.org/10.1088/2041-8205/797/2/L22}.
\adsurl{2014ApJ...797L..22K}.
\end{barticle}
\endbibitem

\bibitem[\protect\citeauthoryear{{Kohutova} and {Verwichte}}{2017}]{koh17}
\begin{barticle}
\bauthor{\bsnm{{Kohutova}}, \binits{P.}},
\bauthor{\bsnm{{Verwichte}}, \binits{E.}}:
\byear{2017},
\batitle{{Excitation of vertical coronal loop oscillations by plasma
  condensations}}.
\bjtitle{\aap}
\bvolume{606},
\bfpage{A120}.
\doiurl{https://doi.org/10.1051/0004-6361/201731417}.
\adsurl{2017A&A...606A.120K}.
\end{barticle}
\endbibitem

\bibitem[\protect\citeauthoryear{{Kumar} et~al.}{2013}]{kum13}
\begin{barticle}
\bauthor{\bsnm{{Kumar}}, \binits{P.}},
\bauthor{\bsnm{{Cho}}, \binits{K.-S.}},
\bauthor{\bsnm{{Chen}}, \binits{P.F.}},
\bauthor{\bsnm{{Bong}}, \binits{S.-C.}},
\bauthor{\bsnm{{Park}}, \binits{S.-H.}}:
\byear{2013},
\batitle{{Multiwavelength Study of a Solar Eruption from AR NOAA 11112: II.
  Large-Scale Coronal Wave and Loop Oscillation}}.
\bjtitle{\solphys}
\bvolume{282},
\bfpage{523}.
\doiurl{https://doi.org/10.1007/s11207-012-0158-7}.
\adsurl{2013SoPh..282..523K}.
\end{barticle}
\endbibitem

\bibitem[\protect\citeauthoryear{{Lemen} et~al.}{2012}]{lem12}
\begin{barticle}
\bauthor{\bsnm{{Lemen}}, \binits{J.R.}},
\bauthor{\bsnm{{Title}}, \binits{A.M.}},
\bauthor{\bsnm{{Akin}}, \binits{D.J.}},
\bauthor{\bsnm{{Boerner}}, \binits{P.F.}},
\bauthor{\bsnm{{Chou}}, \binits{C.}},
\bauthor{\bsnm{{Drake}}, \binits{J.F.}},
\bauthor{\bsnm{{Duncan}}, \binits{D.W.}},
\bauthor{\bsnm{{Edwards}}, \binits{C.G.}},
\bauthor{\bsnm{{Friedlaender}}, \binits{F.M.}},
\bauthor{\bsnm{{Heyman}}, \binits{G.F.}},
\bauthor{\bsnm{{Hurlburt}}, \binits{N.E.}},
\bauthor{\bsnm{{Katz}}, \binits{N.L.}},
\bauthor{\bsnm{{Kushner}}, \binits{G.D.}},
\bauthor{\bsnm{{Levay}}, \binits{M.}},
\bauthor{\bsnm{{Lindgren}}, \binits{R.W.}},
\bauthor{\bsnm{{Mathur}}, \binits{D.P.}},
\bauthor{\bsnm{{McFeaters}}, \binits{E.L.}},
\bauthor{\bsnm{{Mitchell}}, \binits{S.}},
\bauthor{\bsnm{{Rehse}}, \binits{R.A.}},
\bauthor{\bsnm{{Schrijver}}, \binits{C.J.}},
\bauthor{\bsnm{{Springer}}, \binits{L.A.}},
\bauthor{\bsnm{{Stern}}, \binits{R.A.}},
\bauthor{\bsnm{{Tarbell}}, \binits{T.D.}},
\bauthor{\bsnm{{Wuelser}}, \binits{J.-P.}},
\bauthor{\bsnm{{Wolfson}}, \binits{C.J.}},
\bauthor{\bsnm{{Yanari}}, \binits{C.}},
\bauthor{\bsnm{{Bookbinder}}, \binits{J.A.}},
\bauthor{\bsnm{{Cheimets}}, \binits{P.N.}},
\bauthor{\bsnm{{Caldwell}}, \binits{D.}},
\bauthor{\bsnm{{Deluca}}, \binits{E.E.}},
\bauthor{\bsnm{{Gates}}, \binits{R.}},
\bauthor{\bsnm{{Golub}}, \binits{L.}},
\bauthor{\bsnm{{Park}}, \binits{S.}},
\bauthor{\bsnm{{Podgorski}}, \binits{W.A.}},
\bauthor{\bsnm{{Bush}}, \binits{R.I.}},
\bauthor{\bsnm{{Scherrer}}, \binits{P.H.}},
\bauthor{\bsnm{{Gummin}}, \binits{M.A.}},
\bauthor{\bsnm{{Smith}}, \binits{P.}},
\bauthor{\bsnm{{Auker}}, \binits{G.}},
\bauthor{\bsnm{{Jerram}}, \binits{P.}},
\bauthor{\bsnm{{Pool}}, \binits{P.}},
\bauthor{\bsnm{{Soufli}}, \binits{R.}},
\bauthor{\bsnm{{Windt}}, \binits{D.L.}},
\bauthor{\bsnm{{Beardsley}}, \binits{S.}},
\bauthor{\bsnm{{Clapp}}, \binits{M.}},
\bauthor{\bsnm{{Lang}}, \binits{J.}},
\bauthor{\bsnm{{Waltham}}, \binits{N.}}:
\byear{2012},
\batitle{{The Atmospheric Imaging Assembly (AIA) on the Solar Dynamics
  Observatory (SDO)}}.
\bjtitle{\solphys}
\bvolume{275},
\bfpage{17}.
\doiurl{https://doi.org/10.1007/s11207-011-9776-8}.
\adsurl{2012SoPh..275...17L}.
\end{barticle}
\endbibitem

\bibitem[\protect\citeauthoryear{{Li} and {Long}}{2023}]{lid23}
\begin{barticle}
\bauthor{\bsnm{{Li}}, \binits{D.}},
\bauthor{\bsnm{{Long}}, \binits{D.M.}}:
\byear{2023},
\batitle{{A Statistical Study of Short-period Decayless Oscillations of Coronal
  Loops in an Active Region}}.
\bjtitle{\apj}
\bvolume{944},
\bfpage{8}.
\doiurl{https://doi.org/10.3847/1538-4357/acacf4}.
\adsurl{2023ApJ...944....8L}.
\end{barticle}
\endbibitem

\bibitem[\protect\citeauthoryear{{Li} et~al.}{2017}]{lid17}
\begin{barticle}
\bauthor{\bsnm{{Li}}, \binits{D.}},
\bauthor{\bsnm{{Ning}}, \binits{Z.J.}},
\bauthor{\bsnm{{Huang}}, \binits{Y.}},
\bauthor{\bsnm{{Chen}}, \binits{N.-H.}},
\bauthor{\bsnm{{Zhang}}, \binits{Q.M.}},
\bauthor{\bsnm{{Su}}, \binits{Y.N.}},
\bauthor{\bsnm{{Su}}, \binits{W.}}:
\byear{2017},
\batitle{{Doppler Shift Oscillations from a Hot Line Observed by IRIS}}.
\bjtitle{\apj}
\bvolume{849},
\bfpage{113}.
\doiurl{https://doi.org/10.3847/1538-4357/aa9073}.
\adsurl{2017ApJ...849..113L}.
\end{barticle}
\endbibitem

\bibitem[\protect\citeauthoryear{{Liu} et~al.}{2023}]{liu23}
\begin{barticle}
\bauthor{\bsnm{{Liu}}, \binits{J.}},
\bauthor{\bsnm{{Song}}, \binits{A.}},
\bauthor{\bsnm{{Jess}}, \binits{D.B.}},
\bauthor{\bsnm{{Zhang}}, \binits{J.}},
\bauthor{\bsnm{{Mathioudakis}}, \binits{M.}},
\bauthor{\bsnm{{So{\'o}s}}, \binits{S.}},
\bauthor{\bsnm{{Keenan}}, \binits{F.P.}},
\bauthor{\bsnm{{Wang}}, \binits{Y.}},
\bauthor{\bsnm{{Erd{\'e}lyi}}, \binits{R.}}:
\byear{2023},
\batitle{{Power-law Distribution of Solar Cycle-modulated Coronal Jets}}.
\bjtitle{\apjs}
\bvolume{266},
\bfpage{17}.
\doiurl{https://doi.org/10.3847/1538-4365/acc85a}.
\adsurl{2023ApJS..266...17L}.
\end{barticle}
\endbibitem

\bibitem[\protect\citeauthoryear{{Liu} et~al.}{2011}]{liu11}
\begin{barticle}
\bauthor{\bsnm{{Liu}}, \binits{W.}},
\bauthor{\bsnm{{Berger}}, \binits{T.E.}},
\bauthor{\bsnm{{Title}}, \binits{A.M.}},
\bauthor{\bsnm{{Tarbell}}, \binits{T.D.}},
\bauthor{\bsnm{{Low}}, \binits{B.C.}}:
\byear{2011},
\batitle{{Chromospheric Jet and Growing ``Loop'' Observed by Hinode: New
  Evidence of Fan-spine Magnetic Topology Resulting from Flux Emergence}}.
\bjtitle{\apj}
\bvolume{728},
\bfpage{103}.
\doiurl{https://doi.org/10.1088/0004-637X/728/2/103}.
\adsurl{2011ApJ...728..103L}.
\end{barticle}
\endbibitem

\bibitem[\protect\citeauthoryear{{Luna} et~al.}{2014}]{luna14}
\begin{barticle}
\bauthor{\bsnm{{Luna}}, \binits{M.}},
\bauthor{\bsnm{{Knizhnik}}, \binits{K.}},
\bauthor{\bsnm{{Muglach}}, \binits{K.}},
\bauthor{\bsnm{{Karpen}}, \binits{J.}},
\bauthor{\bsnm{{Gilbert}}, \binits{H.}},
\bauthor{\bsnm{{Kucera}}, \binits{T.A.}},
\bauthor{\bsnm{{Uritsky}}, \binits{V.}}:
\byear{2014},
\batitle{{Observations and Implications of Large-amplitude Longitudinal
  Oscillations in a Solar Filament}}.
\bjtitle{\apj}
\bvolume{785},
\bfpage{79}.
\doiurl{https://doi.org/10.1088/0004-637X/785/1/79}.
\adsurl{2014ApJ...785...79L}.
\end{barticle}
\endbibitem

\bibitem[\protect\citeauthoryear{{Mandal} et~al.}{2022a}]{man22a}
\begin{barticle}
\bauthor{\bsnm{{Mandal}}, \binits{S.}},
\bauthor{\bsnm{{Chitta}}, \binits{L.P.}},
\bauthor{\bsnm{{Peter}}, \binits{H.}},
\bauthor{\bsnm{{Solanki}}, \binits{S.K.}},
\bauthor{\bsnm{{Cuadrado}}, \binits{R.A.}},
\bauthor{\bsnm{{Teriaca}}, \binits{L.}},
\bauthor{\bsnm{{Sch{\"u}hle}}, \binits{U.}},
\bauthor{\bsnm{{Berghmans}}, \binits{D.}},
\bauthor{\bsnm{{Auch{\`e}re}}, \binits{F.}}:
\byear{2022}a,
\batitle{{A highly dynamic small-scale jet in a polar coronal hole}}.
\bjtitle{\aap}
\bvolume{664},
\bfpage{A28}.
\doiurl{https://doi.org/10.1051/0004-6361/202243765}.
\adsurl{2022A&A...664A..28M}.
\end{barticle}
\endbibitem

\bibitem[\protect\citeauthoryear{{Mandal} et~al.}{2022b}]{man22b}
\begin{barticle}
\bauthor{\bsnm{{Mandal}}, \binits{S.}},
\bauthor{\bsnm{{Chitta}}, \binits{L.P.}},
\bauthor{\bsnm{{Antolin}}, \binits{P.}},
\bauthor{\bsnm{{Peter}}, \binits{H.}},
\bauthor{\bsnm{{Solanki}}, \binits{S.K.}},
\bauthor{\bsnm{{Auch{\`e}re}}, \binits{F.}},
\bauthor{\bsnm{{Berghmans}}, \binits{D.}},
\bauthor{\bsnm{{Zhukov}}, \binits{A.N.}},
\bauthor{\bsnm{{Teriaca}}, \binits{L.}},
\bauthor{\bsnm{{Cuadrado}}, \binits{R.A.}},
\bauthor{\bsnm{{Sch{\"u}hle}}, \binits{U.}},
\bauthor{\bsnm{{Parenti}}, \binits{S.}},
\bauthor{\bsnm{{Buchlin}}, \binits{{\'E}.}},
\bauthor{\bsnm{{Harra}}, \binits{L.}},
\bauthor{\bsnm{{Verbeeck}}, \binits{C.}},
\bauthor{\bsnm{{Kraaikamp}}, \binits{E.}},
\bauthor{\bsnm{{Long}}, \binits{D.M.}},
\bauthor{\bsnm{{Rodriguez}}, \binits{L.}},
\bauthor{\bsnm{{Pelouze}}, \binits{G.}},
\bauthor{\bsnm{{Schwanitz}}, \binits{C.}},
\bauthor{\bsnm{{Barczynski}}, \binits{K.}},
\bauthor{\bsnm{{Smith}}, \binits{P.J.}}:
\byear{2022}b,
\batitle{{What drives decayless kink oscillations in active-region coronal
  loops on the Sun?}}
\bjtitle{\aap}
\bvolume{666},
\bfpage{L2}.
\doiurl{https://doi.org/10.1051/0004-6361/202244403}.
\adsurl{2022A&A...666L...2M}.
\end{barticle}
\endbibitem

\bibitem[\protect\citeauthoryear{{Mart{\'\i}nez-Sykora} et~al.}{2017}]{mar17}
\begin{barticle}
\bauthor{\bsnm{{Mart{\'\i}nez-Sykora}}, \binits{J.}},
\bauthor{\bsnm{{De Pontieu}}, \binits{B.}},
\bauthor{\bsnm{{Hansteen}}, \binits{V.H.}},
\bauthor{\bsnm{{Rouppe van der Voort}}, \binits{L.}},
\bauthor{\bsnm{{Carlsson}}, \binits{M.}},
\bauthor{\bsnm{{Pereira}}, \binits{T.M.D.}}:
\byear{2017},
\batitle{{On the generation of solar spicules and Alfv{\'e}nic waves}}.
\bjtitle{Science}
\bvolume{356},
\bfpage{1269}.
\doiurl{https://doi.org/10.1126/science.aah5412}.
\adsurl{2017Sci...356.1269M}.
\end{barticle}
\endbibitem

\bibitem[\protect\citeauthoryear{{Moore} et~al.}{2010}]{mo10}
\begin{barticle}
\bauthor{\bsnm{{Moore}}, \binits{R.L.}},
\bauthor{\bsnm{{Cirtain}}, \binits{J.W.}},
\bauthor{\bsnm{{Sterling}}, \binits{A.C.}},
\bauthor{\bsnm{{Falconer}}, \binits{D.A.}}:
\byear{2010},
\batitle{{Dichotomy of Solar Coronal Jets: Standard Jets and Blowout Jets}}.
\bjtitle{\apj}
\bvolume{720},
\bfpage{757}.
\doiurl{https://doi.org/10.1088/0004-637X/720/1/757}.
\adsurl{2010ApJ...720..757M}.
\end{barticle}
\endbibitem

\bibitem[\protect\citeauthoryear{{Moore} et~al.}{2013}]{mo13}
\begin{barticle}
\bauthor{\bsnm{{Moore}}, \binits{R.L.}},
\bauthor{\bsnm{{Sterling}}, \binits{A.C.}},
\bauthor{\bsnm{{Falconer}}, \binits{D.A.}},
\bauthor{\bsnm{{Robe}}, \binits{D.}}:
\byear{2013},
\batitle{{The Cool Component and the Dichotomy, Lateral Expansion, and Axial
  Rotation of Solar X-Ray Jets}}.
\bjtitle{\apj}
\bvolume{769},
\bfpage{134}.
\doiurl{https://doi.org/10.1088/0004-637X/769/2/134}.
\adsurl{2013ApJ...769..134M}.
\end{barticle}
\endbibitem

\bibitem[\protect\citeauthoryear{{Moreno-Insertis}, {Galsgaard}, and
  {Ugarte-Urra}}{2008}]{mo08}
\begin{barticle}
\bauthor{\bsnm{{Moreno-Insertis}}, \binits{F.}},
\bauthor{\bsnm{{Galsgaard}}, \binits{K.}},
\bauthor{\bsnm{{Ugarte-Urra}}, \binits{I.}}:
\byear{2008},
\batitle{{Jets in Coronal Holes: Hinode Observations and Three-dimensional
  Computer Modeling}}.
\bjtitle{\apjl}
\bvolume{673},
\bfpage{L211}.
\doiurl{https://doi.org/10.1086/527560}.
\adsurl{2008ApJ...673L.211M}.
\end{barticle}
\endbibitem

\bibitem[\protect\citeauthoryear{{Mrozek}}{2011}]{mro11}
\begin{barticle}
\bauthor{\bsnm{{Mrozek}}, \binits{T.}}:
\byear{2011},
\batitle{{Failed Eruption of a Filament as a Driver for Vertical Oscillations
  of Coronal Loops}}.
\bjtitle{\solphys}
\bvolume{270},
\bfpage{191}.
\doiurl{https://doi.org/10.1007/s11207-011-9750-5}.
\adsurl{2011SoPh..270..191M}.
\end{barticle}
\endbibitem

\bibitem[\protect\citeauthoryear{{Mulay} et~al.}{2016}]{mu16}
\begin{barticle}
\bauthor{\bsnm{{Mulay}}, \binits{S.M.}},
\bauthor{\bsnm{{Tripathi}}, \binits{D.}},
\bauthor{\bsnm{{Del Zanna}}, \binits{G.}},
\bauthor{\bsnm{{Mason}}, \binits{H.}}:
\byear{2016},
\batitle{{Multiwavelength study of 20 jets that emanate from the periphery of
  active regions}}.
\bjtitle{\aap}
\bvolume{589},
\bfpage{A79}.
\doiurl{https://doi.org/10.1051/0004-6361/201527473}.
\adsurl{2016A&A...589A..79M}.
\end{barticle}
\endbibitem

\bibitem[\protect\citeauthoryear{{Nakariakov} and {Ofman}}{2001}]{naka01}
\begin{barticle}
\bauthor{\bsnm{{Nakariakov}}, \binits{V.M.}},
\bauthor{\bsnm{{Ofman}}, \binits{L.}}:
\byear{2001},
\batitle{{Determination of the coronal magnetic field by coronal loop
  oscillations}}.
\bjtitle{\aap}
\bvolume{372},
\bfpage{L53}.
\doiurl{https://doi.org/10.1051/0004-6361:20010607}.
\adsurl{2001A&A...372L..53N}.
\end{barticle}
\endbibitem

\bibitem[\protect\citeauthoryear{{Nakariakov} et~al.}{1999}]{naka99}
\begin{barticle}
\bauthor{\bsnm{{Nakariakov}}, \binits{V.M.}},
\bauthor{\bsnm{{Ofman}}, \binits{L.}},
\bauthor{\bsnm{{Deluca}}, \binits{E.E.}},
\bauthor{\bsnm{{Roberts}}, \binits{B.}},
\bauthor{\bsnm{{Davila}}, \binits{J.M.}}:
\byear{1999},
\batitle{{TRACE observation of damped coronal loop oscillations: Implications
  for coronal heating}}.
\bjtitle{Science}
\bvolume{285},
\bfpage{862}.
\doiurl{https://doi.org/10.1126/science.285.5429.862}.
\adsurl{1999Sci...285..862N}.
\end{barticle}
\endbibitem

\bibitem[\protect\citeauthoryear{{Nakariakov} et~al.}{2021}]{naka21}
\begin{barticle}
\bauthor{\bsnm{{Nakariakov}}, \binits{V.M.}},
\bauthor{\bsnm{{Anfinogentov}}, \binits{S.A.}},
\bauthor{\bsnm{{Antolin}}, \binits{P.}},
\bauthor{\bsnm{{Jain}}, \binits{R.}},
\bauthor{\bsnm{{Kolotkov}}, \binits{D.Y.}},
\bauthor{\bsnm{{Kupriyanova}}, \binits{E.G.}},
\bauthor{\bsnm{{Li}}, \binits{D.}},
\bauthor{\bsnm{{Magyar}}, \binits{N.}},
\bauthor{\bsnm{{Nistic{\`o}}}, \binits{G.}},
\bauthor{\bsnm{{Pascoe}}, \binits{D.J.}},
\bauthor{\bsnm{{Srivastava}}, \binits{A.K.}},
\bauthor{\bsnm{{Terradas}}, \binits{J.}},
\bauthor{\bsnm{{Vasheghani Farahani}}, \binits{S.}},
\bauthor{\bsnm{{Verth}}, \binits{G.}},
\bauthor{\bsnm{{Yuan}}, \binits{D.}},
\bauthor{\bsnm{{Zimovets}}, \binits{I.V.}}:
\byear{2021},
\batitle{{Kink Oscillations of Coronal Loops}}.
\bjtitle{\ssr}
\bvolume{217},
\bfpage{73}.
\doiurl{https://doi.org/10.1007/s11214-021-00847-2}.
\adsurl{2021SSRv..217...73N}.
\end{barticle}
\endbibitem

\bibitem[\protect\citeauthoryear{{Ni} et~al.}{2017}]{ni17}
\begin{barticle}
\bauthor{\bsnm{{Ni}}, \binits{L.}},
\bauthor{\bsnm{{Zhang}}, \binits{Q.-M.}},
\bauthor{\bsnm{{Murphy}}, \binits{N.A.}},
\bauthor{\bsnm{{Lin}}, \binits{J.}}:
\byear{2017},
\batitle{{Blob Formation and Ejection in Coronal Jets due to the Plasmoid and
  Kelvin-Helmholtz Instabilities}}.
\bjtitle{\apj}
\bvolume{841},
\bfpage{27}.
\doiurl{https://doi.org/10.3847/1538-4357/aa6ffe}.
\adsurl{2017ApJ...841...27N}.
\end{barticle}
\endbibitem

\bibitem[\protect\citeauthoryear{{Ni} et~al.}{2022}]{ni22}
\begin{barticle}
\bauthor{\bsnm{{Ni}}, \binits{Y.W.}},
\bauthor{\bsnm{{Guo}}, \binits{J.H.}},
\bauthor{\bsnm{{Zhang}}, \binits{Q.M.}},
\bauthor{\bsnm{{Chen}}, \binits{J.L.}},
\bauthor{\bsnm{{Fang}}, \binits{C.}},
\bauthor{\bsnm{{Chen}}, \binits{P.F.}}:
\byear{2022},
\batitle{{Decayless longitudinal oscillations of a solar filament maintained by
  quasi-periodic jets}}.
\bjtitle{\aap}
\bvolume{663},
\bfpage{A31}.
\doiurl{https://doi.org/10.1051/0004-6361/202142979}.
\adsurl{2022A&A...663A..31N}.
\end{barticle}
\endbibitem

\bibitem[\protect\citeauthoryear{{Nishizuka} et~al.}{2008}]{nis08}
\begin{barticle}
\bauthor{\bsnm{{Nishizuka}}, \binits{N.}},
\bauthor{\bsnm{{Shimizu}}, \binits{M.}},
\bauthor{\bsnm{{Nakamura}}, \binits{T.}},
\bauthor{\bsnm{{Otsuji}}, \binits{K.}},
\bauthor{\bsnm{{Okamoto}}, \binits{T.J.}},
\bauthor{\bsnm{{Katsukawa}}, \binits{Y.}},
\bauthor{\bsnm{{Shibata}}, \binits{K.}}:
\byear{2008},
\batitle{{Giant Chromospheric Anemone Jet Observed with Hinode and Comparison
  with Magnetohydrodynamic Simulations: Evidence of Propagating Alfv{\'e}n
  Waves and Magnetic Reconnection}}.
\bjtitle{\apjl}
\bvolume{683},
\bfpage{L83}.
\doiurl{https://doi.org/10.1086/591445}.
\adsurl{2008ApJ...683L..83N}.
\end{barticle}
\endbibitem

\bibitem[\protect\citeauthoryear{{Nistic{\`o}}, {Nakariakov}, and
  {Verwichte}}{2013}]{nis13}
\begin{barticle}
\bauthor{\bsnm{{Nistic{\`o}}}, \binits{G.}},
\bauthor{\bsnm{{Nakariakov}}, \binits{V.M.}},
\bauthor{\bsnm{{Verwichte}}, \binits{E.}}:
\byear{2013},
\batitle{{Decaying and decayless transverse oscillations of a coronal loop}}.
\bjtitle{\aap}
\bvolume{552},
\bfpage{A57}.
\doiurl{https://doi.org/10.1051/0004-6361/201220676}.
\adsurl{2013A&A...552A..57N}.
\end{barticle}
\endbibitem

\bibitem[\protect\citeauthoryear{{Nistic{\`o}} et~al.}{2009}]{nis09}
\begin{barticle}
\bauthor{\bsnm{{Nistic{\`o}}}, \binits{G.}},
\bauthor{\bsnm{{Bothmer}}, \binits{V.}},
\bauthor{\bsnm{{Patsourakos}}, \binits{S.}},
\bauthor{\bsnm{{Zimbardo}}, \binits{G.}}:
\byear{2009},
\batitle{{Characteristics of EUV Coronal Jets Observed with STEREO/SECCHI}}.
\bjtitle{\solphys}
\bvolume{259},
\bfpage{87}.
\doiurl{https://doi.org/10.1007/s11207-009-9424-8}.
\adsurl{2009SoPh..259...87N}.
\end{barticle}
\endbibitem

\bibitem[\protect\citeauthoryear{{Nistic{\`o}} et~al.}{2017}]{nis17}
\begin{barticle}
\bauthor{\bsnm{{Nistic{\`o}}}, \binits{G.}},
\bauthor{\bsnm{{Polito}}, \binits{V.}},
\bauthor{\bsnm{{Nakariakov}}, \binits{V.M.}},
\bauthor{\bsnm{{Del Zanna}}, \binits{G.}}:
\byear{2017},
\batitle{{Multi-instrument observations of a failed flare eruption associated
  with MHD waves in a loop bundle}}.
\bjtitle{\aap}
\bvolume{600},
\bfpage{A37}.
\doiurl{https://doi.org/10.1051/0004-6361/201629324}.
\adsurl{2017A&A...600A..37N}.
\end{barticle}
\endbibitem

\bibitem[\protect\citeauthoryear{{N{\'o}brega-Siverio} and
  {Moreno-Insertis}}{2022}]{nob22}
\begin{barticle}
\bauthor{\bsnm{{N{\'o}brega-Siverio}}, \binits{D.}},
\bauthor{\bsnm{{Moreno-Insertis}}, \binits{F.}}:
\byear{2022},
\batitle{{A 2D Model for Coronal Bright Points: Association with Spicules, UV
  Bursts, Surges, and EUV Coronal Jets}}.
\bjtitle{\apjl}
\bvolume{935},
\bfpage{L21}.
\doiurl{https://doi.org/10.3847/2041-8213/ac85b6}.
\adsurl{2022ApJ...935L..21N}.
\end{barticle}
\endbibitem

\bibitem[\protect\citeauthoryear{{Panesar} et~al.}{2016}]{pan16}
\begin{barticle}
\bauthor{\bsnm{{Panesar}}, \binits{N.K.}},
\bauthor{\bsnm{{Sterling}}, \binits{A.C.}},
\bauthor{\bsnm{{Moore}}, \binits{R.L.}},
\bauthor{\bsnm{{Chakrapani}}, \binits{P.}}:
\byear{2016},
\batitle{{Magnetic Flux Cancelation as the Trigger of Solar Quiet-region
  Coronal Jets}}.
\bjtitle{\apjl}
\bvolume{832},
\bfpage{L7}.
\doiurl{https://doi.org/10.3847/2041-8205/832/1/L7}.
\adsurl{2016ApJ...832L...7P}.
\end{barticle}
\endbibitem

\bibitem[\protect\citeauthoryear{{Pariat}, {Antiochos}, and
  {DeVore}}{2009}]{par09}
\begin{barticle}
\bauthor{\bsnm{{Pariat}}, \binits{E.}},
\bauthor{\bsnm{{Antiochos}}, \binits{S.K.}},
\bauthor{\bsnm{{DeVore}}, \binits{C.R.}}:
\byear{2009},
\batitle{{A Model for Solar Polar Jets}}.
\bjtitle{\apj}
\bvolume{691},
\bfpage{61}.
\doiurl{https://doi.org/10.1088/0004-637X/691/1/61}.
\adsurl{2009ApJ...691...61P}.
\end{barticle}
\endbibitem

\bibitem[\protect\citeauthoryear{{Pucci} et~al.}{2013}]{pu13}
\begin{barticle}
\bauthor{\bsnm{{Pucci}}, \binits{S.}},
\bauthor{\bsnm{{Poletto}}, \binits{G.}},
\bauthor{\bsnm{{Sterling}}, \binits{A.C.}},
\bauthor{\bsnm{{Romoli}}, \binits{M.}}:
\byear{2013},
\batitle{{Physical Parameters of Standard and Blowout Jets}}.
\bjtitle{\apj}
\bvolume{776},
\bfpage{16}.
\doiurl{https://doi.org/10.1088/0004-637X/776/1/16}.
\adsurl{2013ApJ...776...16P}.
\end{barticle}
\endbibitem

\bibitem[\protect\citeauthoryear{{Raouafi} et~al.}{2016}]{Raouafi2016}
\begin{barticle}
\bauthor{\bsnm{{Raouafi}}, \binits{N.E.}},
\bauthor{\bsnm{{Patsourakos}}, \binits{S.}},
\bauthor{\bsnm{{Pariat}}, \binits{E.}},
\bauthor{\bsnm{{Young}}, \binits{P.R.}},
\bauthor{\bsnm{{Sterling}}, \binits{A.C.}},
\bauthor{\bsnm{{Savcheva}}, \binits{A.}},
\bauthor{\bsnm{{Shimojo}}, \binits{M.}},
\bauthor{\bsnm{{Moreno-Insertis}}, \binits{F.}},
\bauthor{\bsnm{{DeVore}}, \binits{C.R.}},
\bauthor{\bsnm{{Archontis}}, \binits{V.}},
\bauthor{\bsnm{{T{\"o}r{\"o}k}}, \binits{T.}},
\bauthor{\bsnm{{Mason}}, \binits{H.}},
\bauthor{\bsnm{{Curdt}}, \binits{W.}},
\bauthor{\bsnm{{Meyer}}, \binits{K.}},
\bauthor{\bsnm{{Dalmasse}}, \binits{K.}},
\bauthor{\bsnm{{Matsui}}, \binits{Y.}}:
\byear{2016},
\batitle{{Solar Coronal Jets: Observations, Theory, and Modeling}}.
\bjtitle{\ssr}
\bvolume{201},
\bfpage{1}.
\doiurl{https://doi.org/10.1007/s11214-016-0260-5}.
\adsurl{2016SSRv..201....1R}.
\end{barticle}
\endbibitem

\bibitem[\protect\citeauthoryear{{Reeves} et~al.}{2020}]{ree20}
\begin{barticle}
\bauthor{\bsnm{{Reeves}}, \binits{K.K.}},
\bauthor{\bsnm{{Polito}}, \binits{V.}},
\bauthor{\bsnm{{Chen}}, \binits{B.}},
\bauthor{\bsnm{{Galan}}, \binits{G.}},
\bauthor{\bsnm{{Yu}}, \binits{S.}},
\bauthor{\bsnm{{Liu}}, \binits{W.}},
\bauthor{\bsnm{{Li}}, \binits{G.}}:
\byear{2020},
\batitle{{Hot Plasma Flows and Oscillations in the Loop-top Region During the
  2017 September 10 X8.2 Solar Flare}}.
\bjtitle{\apj}
\bvolume{905},
\bfpage{165}.
\doiurl{https://doi.org/10.3847/1538-4357/abc4e0}.
\adsurl{2020ApJ...905..165R}.
\end{barticle}
\endbibitem

\bibitem[\protect\citeauthoryear{{Roy}}{1973}]{roy73}
\begin{barticle}
\bauthor{\bsnm{{Roy}}, \binits{J.R.}}:
\byear{1973},
\batitle{{The Magnetic Properties of Solar Surges}}.
\bjtitle{\solphys}
\bvolume{28},
\bfpage{95}.
\doiurl{https://doi.org/10.1007/BF00152915}.
\adsurl{1973SoPh...28...95R}.
\end{barticle}
\endbibitem

\bibitem[\protect\citeauthoryear{{Russell}, {Sim{\~o}es}, and
  {Fletcher}}{2015}]{rus15}
\begin{barticle}
\bauthor{\bsnm{{Russell}}, \binits{A.J.B.}},
\bauthor{\bsnm{{Sim{\~o}es}}, \binits{P.J.A.}},
\bauthor{\bsnm{{Fletcher}}, \binits{L.}}:
\byear{2015},
\batitle{{A unified view of coronal loop contraction and oscillation in
  flares}}.
\bjtitle{\aap}
\bvolume{581},
\bfpage{A8}.
\doiurl{https://doi.org/10.1051/0004-6361/201525746}.
\adsurl{2015A&A...581A...8R}.
\end{barticle}
\endbibitem

\bibitem[\protect\citeauthoryear{{Samanta} et~al.}{2019}]{san19}
\begin{barticle}
\bauthor{\bsnm{{Samanta}}, \binits{T.}},
\bauthor{\bsnm{{Tian}}, \binits{H.}},
\bauthor{\bsnm{{Yurchyshyn}}, \binits{V.}},
\bauthor{\bsnm{{Peter}}, \binits{H.}},
\bauthor{\bsnm{{Cao}}, \binits{W.}},
\bauthor{\bsnm{{Sterling}}, \binits{A.}},
\bauthor{\bsnm{{Erd{\'e}lyi}}, \binits{R.}},
\bauthor{\bsnm{{Ahn}}, \binits{K.}},
\bauthor{\bsnm{{Feng}}, \binits{S.}},
\bauthor{\bsnm{{Utz}}, \binits{D.}},
\bauthor{\bsnm{{Banerjee}}, \binits{D.}},
\bauthor{\bsnm{{Chen}}, \binits{Y.}}:
\byear{2019},
\batitle{{Generation of solar spicules and subsequent atmospheric heating}}.
\bjtitle{Science}
\bvolume{366},
\bfpage{890}.
\doiurl{https://doi.org/10.1126/science.aaw2796}.
\adsurl{2019Sci...366..890S}.
\end{barticle}
\endbibitem

\bibitem[\protect\citeauthoryear{{Sarkar} et~al.}{2016}]{sar16}
\begin{barticle}
\bauthor{\bsnm{{Sarkar}}, \binits{S.}},
\bauthor{\bsnm{{Pant}}, \binits{V.}},
\bauthor{\bsnm{{Srivastava}}, \binits{A.K.}},
\bauthor{\bsnm{{Banerjee}}, \binits{D.}}:
\byear{2016},
\batitle{{Transverse Oscillations in a Coronal Loop Triggered by a Jet}}.
\bjtitle{\solphys}
\bvolume{291},
\bfpage{3269}.
\doiurl{https://doi.org/10.1007/s11207-016-1019-6}.
\adsurl{2016SoPh..291.3269S}.
\end{barticle}
\endbibitem

\bibitem[\protect\citeauthoryear{{Scherrer} et~al.}{2012}]{sch12}
\begin{barticle}
\bauthor{\bsnm{{Scherrer}}, \binits{P.H.}},
\bauthor{\bsnm{{Schou}}, \binits{J.}},
\bauthor{\bsnm{{Bush}}, \binits{R.I.}},
\bauthor{\bsnm{{Kosovichev}}, \binits{A.G.}},
\bauthor{\bsnm{{Bogart}}, \binits{R.S.}},
\bauthor{\bsnm{{Hoeksema}}, \binits{J.T.}},
\bauthor{\bsnm{{Liu}}, \binits{Y.}},
\bauthor{\bsnm{{Duvall}}, \binits{T.L.}},
\bauthor{\bsnm{{Zhao}}, \binits{J.}},
\bauthor{\bsnm{{Title}}, \binits{A.M.}},
\bauthor{\bsnm{{Schrijver}}, \binits{C.J.}},
\bauthor{\bsnm{{Tarbell}}, \binits{T.D.}},
\bauthor{\bsnm{{Tomczyk}}, \binits{S.}}:
\byear{2012},
\batitle{{The Helioseismic and Magnetic Imager (HMI) Investigation for the
  Solar Dynamics Observatory (SDO)}}.
\bjtitle{\solphys}
\bvolume{275},
\bfpage{207}.
\doiurl{https://doi.org/10.1007/s11207-011-9834-2}.
\adsurl{2012SoPh..275..207S}.
\end{barticle}
\endbibitem

\bibitem[\protect\citeauthoryear{{Shen}}{2021}]{Shen2021}
\begin{barticle}
\bauthor{\bsnm{{Shen}}, \binits{Y.}}:
\byear{2021},
\batitle{{Observation and modelling of solar jets}}.
\bjtitle{Proceedings of the Royal Society of London Series A}
\bvolume{477},
\bfpage{217}.
\doiurl{https://doi.org/10.1098/rspa.2020.0217}.
\adsurl{2021RSPSA.47700217S}.
\end{barticle}
\endbibitem

\bibitem[\protect\citeauthoryear{{Shen} and {Liu}}{2012}]{shen12}
\begin{barticle}
\bauthor{\bsnm{{Shen}}, \binits{Y.}},
\bauthor{\bsnm{{Liu}}, \binits{Y.}}:
\byear{2012},
\batitle{{Evidence for the Wave Nature of an Extreme Ultraviolet Wave Observed
  by the Atmospheric Imaging Assembly on Board the Solar Dynamics
  Observatory}}.
\bjtitle{\apj}
\bvolume{754},
\bfpage{7}.
\doiurl{https://doi.org/10.1088/0004-637X/754/1/7}.
\adsurl{2012ApJ...754....7S}.
\end{barticle}
\endbibitem

\bibitem[\protect\citeauthoryear{{Shen} et~al.}{2018}]{shen18}
\begin{barticle}
\bauthor{\bsnm{{Shen}}, \binits{Y.}},
\bauthor{\bsnm{{Tang}}, \binits{Z.}},
\bauthor{\bsnm{{Miao}}, \binits{Y.}},
\bauthor{\bsnm{{Su}}, \binits{J.}},
\bauthor{\bsnm{{Liu}}, \binits{Y.}}:
\byear{2018},
\batitle{{EUV Waves Driven by the Sudden Expansion of Transequatorial Loops
  Caused by Coronal Jets}}.
\bjtitle{\apjl}
\bvolume{860},
\bfpage{L8}.
\doiurl{https://doi.org/10.3847/2041-8213/aac8dd}.
\adsurl{2018ApJ...860L...8S}.
\end{barticle}
\endbibitem

\bibitem[\protect\citeauthoryear{{Shen} et~al.}{2019}]{shen19}
\begin{barticle}
\bauthor{\bsnm{{Shen}}, \binits{Y.}},
\bauthor{\bsnm{{Qu}}, \binits{Z.}},
\bauthor{\bsnm{{Yuan}}, \binits{D.}},
\bauthor{\bsnm{{Chen}}, \binits{H.}},
\bauthor{\bsnm{{Duan}}, \binits{Y.}},
\bauthor{\bsnm{{Zhou}}, \binits{C.}},
\bauthor{\bsnm{{Tang}}, \binits{Z.}},
\bauthor{\bsnm{{Huang}}, \binits{J.}},
\bauthor{\bsnm{{Liu}}, \binits{Y.}}:
\byear{2019},
\batitle{{Stereoscopic Observations of an Erupting Mini-filament-driven
  Two-sided-loop Jet and the Applications for Diagnosing a Filament Magnetic
  Field}}.
\bjtitle{\apj}
\bvolume{883},
\bfpage{104}.
\doiurl{https://doi.org/10.3847/1538-4357/ab3a4d}.
\adsurl{2019ApJ...883..104S}.
\end{barticle}
\endbibitem

\bibitem[\protect\citeauthoryear{{Shi}, {Ning}, and {Li}}{2022}]{shi22}
\begin{barticle}
\bauthor{\bsnm{{Shi}}, \binits{F.}},
\bauthor{\bsnm{{Ning}}, \binits{Z.}},
\bauthor{\bsnm{{Li}}, \binits{D.}}:
\byear{2022},
\batitle{{Investigation of the Oscillations in a Flare-productive Active
  Region}}.
\bjtitle{Research in Astronomy and Astrophysics}
\bvolume{22},
\bfpage{105017}.
\doiurl{https://doi.org/10.1088/1674-4527/ac8f8a}.
\adsurl{2022RAA....22j5017S}.
\end{barticle}
\endbibitem

\bibitem[\protect\citeauthoryear{{Shibata} et~al.}{1992}]{shi92}
\begin{barticle}
\bauthor{\bsnm{{Shibata}}, \binits{K.}},
\bauthor{\bsnm{{Ishido}}, \binits{Y.}},
\bauthor{\bsnm{{Acton}}, \binits{L.W.}},
\bauthor{\bsnm{{Strong}}, \binits{K.T.}},
\bauthor{\bsnm{{Hirayama}}, \binits{T.}},
\bauthor{\bsnm{{Uchida}}, \binits{Y.}},
\bauthor{\bsnm{{McAllister}}, \binits{A.H.}},
\bauthor{\bsnm{{Matsumoto}}, \binits{R.}},
\bauthor{\bsnm{{Tsuneta}}, \binits{S.}},
\bauthor{\bsnm{{Shimizu}}, \binits{T.}},
\bauthor{\bsnm{{Hara}}, \binits{H.}},
\bauthor{\bsnm{{Sakurai}}, \binits{T.}},
\bauthor{\bsnm{{Ichimoto}}, \binits{K.}},
\bauthor{\bsnm{{Nishino}}, \binits{Y.}},
\bauthor{\bsnm{{Ogawara}}, \binits{Y.}}:
\byear{1992},
\batitle{{Observations of X-Ray Jets with the YOHKOH Soft X-Ray Telescope}}.
\bjtitle{\pasj}
\bvolume{44},
\bfpage{L173}.
\adsurl{1992PASJ...44L.173S}.
\end{barticle}
\endbibitem

\bibitem[\protect\citeauthoryear{{Shibata} et~al.}{1994}]{shi94}
\begin{barticle}
\bauthor{\bsnm{{Shibata}}, \binits{K.}},
\bauthor{\bsnm{{Nitta}}, \binits{N.}},
\bauthor{\bsnm{{Strong}}, \binits{K.T.}},
\bauthor{\bsnm{{Matsumoto}}, \binits{R.}},
\bauthor{\bsnm{{Yokoyama}}, \binits{T.}},
\bauthor{\bsnm{{Hirayama}}, \binits{T.}},
\bauthor{\bsnm{{Hudson}}, \binits{H.}},
\bauthor{\bsnm{{Ogawara}}, \binits{Y.}}:
\byear{1994},
\batitle{{A Gigantic Coronal Jet Ejected from a Compact Active Region in a
  Coronal Hole}}.
\bjtitle{\apjl}
\bvolume{431},
\bfpage{L51}.
\doiurl{https://doi.org/10.1086/187470}.
\adsurl{1994ApJ...431L..51S}.
\end{barticle}
\endbibitem

\bibitem[\protect\citeauthoryear{{Shibata} et~al.}{2007}]{shi07}
\begin{barticle}
\bauthor{\bsnm{{Shibata}}, \binits{K.}},
\bauthor{\bsnm{{Nakamura}}, \binits{T.}},
\bauthor{\bsnm{{Matsumoto}}, \binits{T.}},
\bauthor{\bsnm{{Otsuji}}, \binits{K.}},
\bauthor{\bsnm{{Okamoto}}, \binits{T.J.}},
\bauthor{\bsnm{{Nishizuka}}, \binits{N.}},
\bauthor{\bsnm{{Kawate}}, \binits{T.}},
\bauthor{\bsnm{{Watanabe}}, \binits{H.}},
\bauthor{\bsnm{{Nagata}}, \binits{S.}},
\bauthor{\bsnm{{UeNo}}, \binits{S.}},
\bauthor{\bsnm{{Kitai}}, \binits{R.}},
\bauthor{\bsnm{{Nozawa}}, \binits{S.}},
\bauthor{\bsnm{{Tsuneta}}, \binits{S.}},
\bauthor{\bsnm{{Suematsu}}, \binits{Y.}},
\bauthor{\bsnm{{Ichimoto}}, \binits{K.}},
\bauthor{\bsnm{{Shimizu}}, \binits{T.}},
\bauthor{\bsnm{{Katsukawa}}, \binits{Y.}},
\bauthor{\bsnm{{Tarbell}}, \binits{T.D.}},
\bauthor{\bsnm{{Berger}}, \binits{T.E.}},
\bauthor{\bsnm{{Lites}}, \binits{B.W.}},
\bauthor{\bsnm{{Shine}}, \binits{R.A.}},
\bauthor{\bsnm{{Title}}, \binits{A.M.}}:
\byear{2007},
\batitle{{Chromospheric Anemone Jets as Evidence of Ubiquitous Reconnection}}.
\bjtitle{Science}
\bvolume{318},
\bfpage{1591}.
\doiurl{https://doi.org/10.1126/science.1146708}.
\adsurl{2007Sci...318.1591S}.
\end{barticle}
\endbibitem

\bibitem[\protect\citeauthoryear{{Shimojo} and {Shibata}}{2000}]{shi00}
\begin{barticle}
\bauthor{\bsnm{{Shimojo}}, \binits{M.}},
\bauthor{\bsnm{{Shibata}}, \binits{K.}}:
\byear{2000},
\batitle{{Physical Parameters of Solar X-Ray Jets}}.
\bjtitle{\apj}
\bvolume{542},
\bfpage{1100}.
\doiurl{https://doi.org/10.1086/317024}.
\adsurl{2000ApJ...542.1100S}.
\end{barticle}
\endbibitem

\bibitem[\protect\citeauthoryear{{Shimojo} et~al.}{1996}]{shi96}
\begin{barticle}
\bauthor{\bsnm{{Shimojo}}, \binits{M.}},
\bauthor{\bsnm{{Hashimoto}}, \binits{S.}},
\bauthor{\bsnm{{Shibata}}, \binits{K.}},
\bauthor{\bsnm{{Hirayama}}, \binits{T.}},
\bauthor{\bsnm{{Hudson}}, \binits{H.S.}},
\bauthor{\bsnm{{Acton}}, \binits{L.W.}}:
\byear{1996},
\batitle{{Statistical Study of Solar X-Ray Jets Observed with the YOHKOH Soft
  X-Ray Telescope}}.
\bjtitle{\pasj}
\bvolume{48},
\bfpage{123}.
\doiurl{https://doi.org/10.1093/pasj/48.1.123}.
\adsurl{1996PASJ...48..123S}.
\end{barticle}
\endbibitem

\bibitem[\protect\citeauthoryear{{Sim{\~o}es} et~al.}{2013}]{sim13}
\begin{barticle}
\bauthor{\bsnm{{Sim{\~o}es}}, \binits{P.J.A.}},
\bauthor{\bsnm{{Fletcher}}, \binits{L.}},
\bauthor{\bsnm{{Hudson}}, \binits{H.S.}},
\bauthor{\bsnm{{Russell}}, \binits{A.J.B.}}:
\byear{2013},
\batitle{{Implosion of Coronal Loops during the Impulsive Phase of a Solar
  Flare}}.
\bjtitle{\apj}
\bvolume{777},
\bfpage{152}.
\doiurl{https://doi.org/10.1088/0004-637X/777/2/152}.
\adsurl{2013ApJ...777..152S}.
\end{barticle}
\endbibitem

\bibitem[\protect\citeauthoryear{{Singh} et~al.}{2012}]{sin12}
\begin{barticle}
\bauthor{\bsnm{{Singh}}, \binits{K.A.P.}},
\bauthor{\bsnm{{Isobe}}, \binits{H.}},
\bauthor{\bsnm{{Nishizuka}}, \binits{N.}},
\bauthor{\bsnm{{Nishida}}, \binits{K.}},
\bauthor{\bsnm{{Shibata}}, \binits{K.}}:
\byear{2012},
\batitle{{Multiple Plasma Ejections and Intermittent Nature of Magnetic
  Reconnection in Solar Chromospheric Anemone Jets}}.
\bjtitle{\apj}
\bvolume{759},
\bfpage{33}.
\doiurl{https://doi.org/10.1088/0004-637X/759/1/33}.
\adsurl{2012ApJ...759...33S}.
\end{barticle}
\endbibitem

\bibitem[\protect\citeauthoryear{{Srivastava} and {Goossens}}{2013}]{sri13}
\begin{barticle}
\bauthor{\bsnm{{Srivastava}}, \binits{A.K.}},
\bauthor{\bsnm{{Goossens}}, \binits{M.}}:
\byear{2013},
\batitle{{X6.9-class Flare-induced Vertical Kink Oscillations in a Large-scale
  Plasma Curtain as Observed by the Solar Dynamics Observatory/Atmospheric
  Imaging Assembly}}.
\bjtitle{\apj}
\bvolume{777},
\bfpage{17}.
\doiurl{https://doi.org/10.1088/0004-637X/777/1/17}.
\adsurl{2013ApJ...777...17S}.
\end{barticle}
\endbibitem

\bibitem[\protect\citeauthoryear{{Sterling}, {Moore}, and
  {Panesar}}{2022}]{ste22}
\begin{barticle}
\bauthor{\bsnm{{Sterling}}, \binits{A.C.}},
\bauthor{\bsnm{{Moore}}, \binits{R.L.}},
\bauthor{\bsnm{{Panesar}}, \binits{N.K.}}:
\byear{2022},
\batitle{{Another Look at Erupting Minifilaments at the Base of Solar X-Ray
  Polar Coronal ``Standard'' and ``Blowout'' Jets}}.
\bjtitle{\apj}
\bvolume{927},
\bfpage{127}.
\doiurl{https://doi.org/10.3847/1538-4357/ac473f}.
\adsurl{2022ApJ...927..127S}.
\end{barticle}
\endbibitem

\bibitem[\protect\citeauthoryear{{Sterling} et~al.}{2015}]{ste15}
\begin{barticle}
\bauthor{\bsnm{{Sterling}}, \binits{A.C.}},
\bauthor{\bsnm{{Moore}}, \binits{R.L.}},
\bauthor{\bsnm{{Falconer}}, \binits{D.A.}},
\bauthor{\bsnm{{Adams}}, \binits{M.}}:
\byear{2015},
\batitle{{Small-scale filament eruptions as the driver of X-ray jets in solar
  coronal holes}}.
\bjtitle{\nat}
\bvolume{523},
\bfpage{437}.
\doiurl{https://doi.org/10.1038/nature14556}.
\adsurl{2015Natur.523..437S}.
\end{barticle}
\endbibitem

\bibitem[\protect\citeauthoryear{{Sterling} et~al.}{2016}]{ste16}
\begin{barticle}
\bauthor{\bsnm{{Sterling}}, \binits{A.C.}},
\bauthor{\bsnm{{Moore}}, \binits{R.L.}},
\bauthor{\bsnm{{Falconer}}, \binits{D.A.}},
\bauthor{\bsnm{{Panesar}}, \binits{N.K.}},
\bauthor{\bsnm{{Akiyama}}, \binits{S.}},
\bauthor{\bsnm{{Yashiro}}, \binits{S.}},
\bauthor{\bsnm{{Gopalswamy}}, \binits{N.}}:
\byear{2016},
\batitle{{Minifilament Eruptions that Drive Coronal Jets in a Solar Active
  Region}}.
\bjtitle{\apj}
\bvolume{821},
\bfpage{100}.
\doiurl{https://doi.org/10.3847/0004-637X/821/2/100}.
\adsurl{2016ApJ...821..100S}.
\end{barticle}
\endbibitem

\bibitem[\protect\citeauthoryear{{Tan} et~al.}{2023}]{tan23}
\begin{barticle}
\bauthor{\bsnm{{Tan}}, \binits{S.}},
\bauthor{\bsnm{{Shen}}, \binits{Y.}},
\bauthor{\bsnm{{Zhou}}, \binits{X.}},
\bauthor{\bsnm{{Tang}}, \binits{Z.}},
\bauthor{\bsnm{{Zhou}}, \binits{C.}},
\bauthor{\bsnm{{Duan}}, \binits{Y.}},
\bauthor{\bsnm{{Yao}}, \binits{S.}}:
\byear{2023},
\batitle{{Stereoscopic observation of simultaneous longitudinal and transverse
  oscillations in a single filament driven by two-sided-loop jet}}.
\bjtitle{\mnras}
\bvolume{520},
\bfpage{3080}.
\doiurl{https://doi.org/10.1093/mnras/stad295}.
\adsurl{2023MNRAS.520.3080T}.
\end{barticle}
\endbibitem

\bibitem[\protect\citeauthoryear{{Tian} et~al.}{2012}]{tian12}
\begin{barticle}
\bauthor{\bsnm{{Tian}}, \binits{H.}},
\bauthor{\bsnm{{McIntosh}}, \binits{S.W.}},
\bauthor{\bsnm{{Wang}}, \binits{T.}},
\bauthor{\bsnm{{Ofman}}, \binits{L.}},
\bauthor{\bsnm{{De Pontieu}}, \binits{B.}},
\bauthor{\bsnm{{Innes}}, \binits{D.E.}},
\bauthor{\bsnm{{Peter}}, \binits{H.}}:
\byear{2012},
\batitle{{Persistent Doppler Shift Oscillations Observed with Hinode/EIS in the
  Solar Corona: Spectroscopic Signatures of Alfv{\'e}nic Waves and Recurring
  Upflows}}.
\bjtitle{\apj}
\bvolume{759},
\bfpage{144}.
\doiurl{https://doi.org/10.1088/0004-637X/759/2/144}.
\adsurl{2012ApJ...759..144T}.
\end{barticle}
\endbibitem

\bibitem[\protect\citeauthoryear{{Tian} et~al.}{2014a}]{tian14a}
\begin{barticle}
\bauthor{\bsnm{{Tian}}, \binits{H.}},
\bauthor{\bsnm{{Li}}, \binits{G.}},
\bauthor{\bsnm{{Reeves}}, \binits{K.K.}},
\bauthor{\bsnm{{Raymond}}, \binits{J.C.}},
\bauthor{\bsnm{{Guo}}, \binits{F.}},
\bauthor{\bsnm{{Liu}}, \binits{W.}},
\bauthor{\bsnm{{Chen}}, \binits{B.}},
\bauthor{\bsnm{{Murphy}}, \binits{N.A.}}:
\byear{2014}a,
\batitle{{Imaging and Spectroscopic Observations of Magnetic Reconnection and
  Chromospheric Evaporation in a Solar Flare}}.
\bjtitle{\apjl}
\bvolume{797},
\bfpage{L14}.
\doiurl{https://doi.org/10.1088/2041-8205/797/2/L14}.
\adsurl{2014ApJ...797L..14T}.
\end{barticle}
\endbibitem

\bibitem[\protect\citeauthoryear{{Tian} et~al.}{2014b}]{tian14b}
\begin{barticle}
\bauthor{\bsnm{{Tian}}, \binits{H.}},
\bauthor{\bsnm{{DeLuca}}, \binits{E.E.}},
\bauthor{\bsnm{{Cranmer}}, \binits{S.R.}},
\bauthor{\bsnm{{De Pontieu}}, \binits{B.}},
\bauthor{\bsnm{{Peter}}, \binits{H.}},
\bauthor{\bsnm{{Mart{\'\i}nez-Sykora}}, \binits{J.}},
\bauthor{\bsnm{{Golub}}, \binits{L.}},
\bauthor{\bsnm{{McKillop}}, \binits{S.}},
\bauthor{\bsnm{{Reeves}}, \binits{K.K.}},
\bauthor{\bsnm{{Miralles}}, \binits{M.P.}},
\bauthor{\bsnm{{McCauley}}, \binits{P.}},
\bauthor{\bsnm{{Saar}}, \binits{S.}},
\bauthor{\bsnm{{Testa}}, \binits{P.}},
\bauthor{\bsnm{{Weber}}, \binits{M.}},
\bauthor{\bsnm{{Murphy}}, \binits{N.}},
\bauthor{\bsnm{{Lemen}}, \binits{J.}},
\bauthor{\bsnm{{Title}}, \binits{A.}},
\bauthor{\bsnm{{Boerner}}, \binits{P.}},
\bauthor{\bsnm{{Hurlburt}}, \binits{N.}},
\bauthor{\bsnm{{Tarbell}}, \binits{T.D.}},
\bauthor{\bsnm{{Wuelser}}, \binits{J.P.}},
\bauthor{\bsnm{{Kleint}}, \binits{L.}},
\bauthor{\bsnm{{Kankelborg}}, \binits{C.}},
\bauthor{\bsnm{{Jaeggli}}, \binits{S.}},
\bauthor{\bsnm{{Carlsson}}, \binits{M.}},
\bauthor{\bsnm{{Hansteen}}, \binits{V.}},
\bauthor{\bsnm{{McIntosh}}, \binits{S.W.}}:
\byear{2014}b,
\batitle{{Prevalence of small-scale jets from the networks of the solar
  transition region and chromosphere}}.
\bjtitle{Science}
\bvolume{346},
\bfpage{1255711}.
\doiurl{https://doi.org/10.1126/science.1255711}.
\adsurl{2014Sci...346A.315T}.
\end{barticle}
\endbibitem

\bibitem[\protect\citeauthoryear{{Van Doorsselaere} et~al.}{2008}]{van08}
\begin{barticle}
\bauthor{\bsnm{{Van Doorsselaere}}, \binits{T.}},
\bauthor{\bsnm{{Nakariakov}}, \binits{V.M.}},
\bauthor{\bsnm{{Young}}, \binits{P.R.}},
\bauthor{\bsnm{{Verwichte}}, \binits{E.}}:
\byear{2008},
\batitle{{Coronal magnetic field measurement using loop oscillations observed
  by Hinode/EIS}}.
\bjtitle{\aap}
\bvolume{487},
\bfpage{L17}.
\doiurl{https://doi.org/10.1051/0004-6361:200810186}.
\adsurl{2008A&A...487L..17V}.
\end{barticle}
\endbibitem

\bibitem[\protect\citeauthoryear{{Verwichte} and {Kohutova}}{2017}]{ver17}
\begin{barticle}
\bauthor{\bsnm{{Verwichte}}, \binits{E.}},
\bauthor{\bsnm{{Kohutova}}, \binits{P.}}:
\byear{2017},
\batitle{{Excitation and evolution of vertically polarised transverse loop
  oscillations by coronal rain}}.
\bjtitle{\aap}
\bvolume{601},
\bfpage{L2}.
\doiurl{https://doi.org/10.1051/0004-6361/201730675}.
\adsurl{2017A&A...601L...2V}.
\end{barticle}
\endbibitem

\bibitem[\protect\citeauthoryear{{Verwichte}, {Foullon}, and {Van
  Doorsselaere}}{2010}]{ver10}
\begin{barticle}
\bauthor{\bsnm{{Verwichte}}, \binits{E.}},
\bauthor{\bsnm{{Foullon}}, \binits{C.}},
\bauthor{\bsnm{{Van Doorsselaere}}, \binits{T.}}:
\byear{2010},
\batitle{{Spatial Seismology of a Large Coronal Loop Arcade from TRACE and EIT
  Observations of its Transverse Oscillations}}.
\bjtitle{\apj}
\bvolume{717},
\bfpage{458}.
\doiurl{https://doi.org/10.1088/0004-637X/717/1/458}.
\adsurl{2010ApJ...717..458V}.
\end{barticle}
\endbibitem

\bibitem[\protect\citeauthoryear{{Verwichte} et~al.}{2009}]{ver09}
\begin{barticle}
\bauthor{\bsnm{{Verwichte}}, \binits{E.}},
\bauthor{\bsnm{{Aschwanden}}, \binits{M.J.}},
\bauthor{\bsnm{{Van Doorsselaere}}, \binits{T.}},
\bauthor{\bsnm{{Foullon}}, \binits{C.}},
\bauthor{\bsnm{{Nakariakov}}, \binits{V.M.}}:
\byear{2009},
\batitle{{Seismology of a Large Solar Coronal Loop from EUVI/STEREO
  Observations of its Transverse Oscillation}}.
\bjtitle{\apj}
\bvolume{698},
\bfpage{397}.
\doiurl{https://doi.org/10.1088/0004-637X/698/1/397}.
\adsurl{2009ApJ...698..397V}.
\end{barticle}
\endbibitem

\bibitem[\protect\citeauthoryear{{Verwichte} et~al.}{2013}]{ver13}
\begin{barticle}
\bauthor{\bsnm{{Verwichte}}, \binits{E.}},
\bauthor{\bsnm{{Van Doorsselaere}}, \binits{T.}},
\bauthor{\bsnm{{White}}, \binits{R.S.}},
\bauthor{\bsnm{{Antolin}}, \binits{P.}}:
\byear{2013},
\batitle{{Statistical seismology of transverse waves in the solar corona}}.
\bjtitle{\aap}
\bvolume{552},
\bfpage{A138}.
\doiurl{https://doi.org/10.1051/0004-6361/201220456}.
\adsurl{2013A&A...552A.138V}.
\end{barticle}
\endbibitem

\bibitem[\protect\citeauthoryear{{Wang} and {Solanki}}{2004}]{wang04}
\begin{barticle}
\bauthor{\bsnm{{Wang}}, \binits{T.J.}},
\bauthor{\bsnm{{Solanki}}, \binits{S.K.}}:
\byear{2004},
\batitle{{Vertical oscillations of a coronal loop observed by TRACE}}.
\bjtitle{\aap}
\bvolume{421},
\bfpage{L33}.
\doiurl{https://doi.org/10.1051/0004-6361:20040186}.
\adsurl{2004A&A...421L..33W}.
\end{barticle}
\endbibitem

\bibitem[\protect\citeauthoryear{{Wang} et~al.}{2021}]{wang21}
\begin{barticle}
\bauthor{\bsnm{{Wang}}, \binits{T.}},
\bauthor{\bsnm{{Ofman}}, \binits{L.}},
\bauthor{\bsnm{{Yuan}}, \binits{D.}},
\bauthor{\bsnm{{Reale}}, \binits{F.}},
\bauthor{\bsnm{{Kolotkov}}, \binits{D.Y.}},
\bauthor{\bsnm{{Srivastava}}, \binits{A.K.}}:
\byear{2021},
\batitle{{Slow-Mode Magnetoacoustic Waves in Coronal Loops}}.
\bjtitle{\ssr}
\bvolume{217},
\bfpage{34}.
\doiurl{https://doi.org/10.1007/s11214-021-00811-0}.
\adsurl{2021SSRv..217...34W}.
\end{barticle}
\endbibitem

\bibitem[\protect\citeauthoryear{{Wang} et~al.}{2023}]{wy23}
\begin{barticle}
\bauthor{\bsnm{{Wang}}, \binits{Y.}},
\bauthor{\bsnm{{Zhang}}, \binits{Q.}},
\bauthor{\bsnm{{Hong}}, \binits{Z.}},
\bauthor{\bsnm{{Shen}}, \binits{J.}},
\bauthor{\bsnm{{Ji}}, \binits{H.}},
\bauthor{\bsnm{{Cao}}, \binits{W.}}:
\byear{2023},
\batitle{{High-resolution He I 10 830 {\r{A}} narrowband imaging for precursors
  of chromospheric jets and their quasi-periodic properties}}.
\bjtitle{\aap}
\bvolume{672},
\bfpage{A173}.
\doiurl{https://doi.org/10.1051/0004-6361/202244607}.
\adsurl{2023A&A...672A.173W}.
\end{barticle}
\endbibitem

\bibitem[\protect\citeauthoryear{{White} and {Verwichte}}{2012}]{wht12a}
\begin{barticle}
\bauthor{\bsnm{{White}}, \binits{R.S.}},
\bauthor{\bsnm{{Verwichte}}, \binits{E.}}:
\byear{2012},
\batitle{{Transverse coronal loop oscillations seen in unprecedented detail by
  AIA/SDO}}.
\bjtitle{\aap}
\bvolume{537},
\bfpage{A49}.
\doiurl{https://doi.org/10.1051/0004-6361/201118093}.
\adsurl{2012A&A...537A..49W}.
\end{barticle}
\endbibitem

\bibitem[\protect\citeauthoryear{{White}, {Verwichte}, and
  {Foullon}}{2012}]{wht12b}
\begin{barticle}
\bauthor{\bsnm{{White}}, \binits{R.S.}},
\bauthor{\bsnm{{Verwichte}}, \binits{E.}},
\bauthor{\bsnm{{Foullon}}, \binits{C.}}:
\byear{2012},
\batitle{{First observation of a transverse vertical oscillation during the
  formation of a hot post-flare loop}}.
\bjtitle{\aap}
\bvolume{545},
\bfpage{A129}.
\doiurl{https://doi.org/10.1051/0004-6361/201219856}.
\adsurl{2012A&A...545A.129W}.
\end{barticle}
\endbibitem

\bibitem[\protect\citeauthoryear{{Wyper}, {DeVore}, and
  {Antiochos}}{2018}]{wp18}
\begin{barticle}
\bauthor{\bsnm{{Wyper}}, \binits{P.F.}},
\bauthor{\bsnm{{DeVore}}, \binits{C.R.}},
\bauthor{\bsnm{{Antiochos}}, \binits{S.K.}}:
\byear{2018},
\batitle{{A Breakout Model for Solar Coronal Jets with Filaments}}.
\bjtitle{\apj}
\bvolume{852},
\bfpage{98}.
\doiurl{https://doi.org/10.3847/1538-4357/aa9ffc}.
\adsurl{2018ApJ...852...98W}.
\end{barticle}
\endbibitem

\bibitem[\protect\citeauthoryear{{Yang} et~al.}{2019}]{yang19}
\begin{barticle}
\bauthor{\bsnm{{Yang}}, \binits{L.}},
\bauthor{\bsnm{{Yan}}, \binits{X.}},
\bauthor{\bsnm{{Xue}}, \binits{Z.}},
\bauthor{\bsnm{{Li}}, \binits{T.}},
\bauthor{\bsnm{{Wang}}, \binits{J.}},
\bauthor{\bsnm{{Li}}, \binits{Q.}},
\bauthor{\bsnm{{Cheng}}, \binits{X.}}:
\byear{2019},
\batitle{{Transfer of Twists from a Mini-filament to Large-scale Loops by
  Magnetic Reconnection}}.
\bjtitle{\apj}
\bvolume{887},
\bfpage{239}.
\doiurl{https://doi.org/10.3847/1538-4357/ab55d7}.
\adsurl{2019ApJ...887..239Y}.
\end{barticle}
\endbibitem

\bibitem[\protect\citeauthoryear{{Yang} et~al.}{2024a}]{ylh24}
\begin{barticle}
\bauthor{\bsnm{{Yang}}, \binits{L.}},
\bauthor{\bsnm{{Yan}}, \binits{X.}},
\bauthor{\bsnm{{Xue}}, \binits{Z.}},
\bauthor{\bsnm{{Xu}}, \binits{Z.}},
\bauthor{\bsnm{{Zhang}}, \binits{Q.}},
\bauthor{\bsnm{{Hou}}, \binits{Y.}},
\bauthor{\bsnm{{Wang}}, \binits{J.}},
\bauthor{\bsnm{{Chen}}, \binits{H.}},
\bauthor{\bsnm{{Li}}, \binits{Q.}}:
\byear{2024}a,
\batitle{{Simultaneous observations of a breakout current sheet and a flare
  current sheet in a coronal jet event}}.
\bjtitle{\mnras}
\bvolume{528},
\bfpage{1094}.
\doiurl{https://doi.org/10.1093/mnras/stad3876}.
\adsurl{2024MNRAS.528.1094Y}.
\end{barticle}
\endbibitem

\bibitem[\protect\citeauthoryear{{Yang} et~al.}{2024b}]{ylp24}
\begin{barticle}
\bauthor{\bsnm{{Yang}}, \binits{L.}},
\bauthor{\bsnm{{Xue}}, \binits{Z.}},
\bauthor{\bsnm{{Wang}}, \binits{J.}},
\bauthor{\bsnm{{Yang}}, \binits{L.}},
\bauthor{\bsnm{{Li}}, \binits{Q.}},
\bauthor{\bsnm{{Zhou}}, \binits{Y.}},
\bauthor{\bsnm{{Peng}}, \binits{Y.}},
\bauthor{\bsnm{{Zhang}}, \binits{X.}}:
\byear{2024}b,
\batitle{{Two Intermittent Eruptions of a Minifilament Triggered by a Two-step
  Magnetic Reconnection Within a Fan-spine Configuration}}.
\bjtitle{\apj}
\bvolume{976},
\bfpage{135}.
\doiurl{https://doi.org/10.3847/1538-4357/ad84f9}.
\adsurl{2024ApJ...976..135Y}.
\end{barticle}
\endbibitem

\bibitem[\protect\citeauthoryear{{Yang} et~al.}{2020}]{yzh20}
\begin{barticle}
\bauthor{\bsnm{{Yang}}, \binits{Z.}},
\bauthor{\bsnm{{Bethge}}, \binits{C.}},
\bauthor{\bsnm{{Tian}}, \binits{H.}},
\bauthor{\bsnm{{Tomczyk}}, \binits{S.}},
\bauthor{\bsnm{{Morton}}, \binits{R.}},
\bauthor{\bsnm{{Del Zanna}}, \binits{G.}},
\bauthor{\bsnm{{McIntosh}}, \binits{S.W.}},
\bauthor{\bsnm{{Karak}}, \binits{B.B.}},
\bauthor{\bsnm{{Gibson}}, \binits{S.}},
\bauthor{\bsnm{{Samanta}}, \binits{T.}},
\bauthor{\bsnm{{He}}, \binits{J.}},
\bauthor{\bsnm{{Chen}}, \binits{Y.}},
\bauthor{\bsnm{{Wang}}, \binits{L.}}:
\byear{2020},
\batitle{{Global maps of the magnetic field in the solar corona}}.
\bjtitle{Science}
\bvolume{369},
\bfpage{694}.
\doiurl{https://doi.org/10.1126/science.abb4462}.
\adsurl{2020Sci...369..694Y}.
\end{barticle}
\endbibitem

\bibitem[\protect\citeauthoryear{{Yang} et~al.}{2024c}]{yzh24}
\begin{barticle}
\bauthor{\bsnm{{Yang}}, \binits{Z.}},
\bauthor{\bsnm{{Tian}}, \binits{H.}},
\bauthor{\bsnm{{Tomczyk}}, \binits{S.}},
\bauthor{\bsnm{{Liu}}, \binits{X.}},
\bauthor{\bsnm{{Gibson}}, \binits{S.}},
\bauthor{\bsnm{{Morton}}, \binits{R.J.}},
\bauthor{\bsnm{{Downs}}, \binits{C.}}:
\byear{2024}c,
\batitle{{Observing the evolution of the Sun's global coronal magnetic field
  over 8 months}}.
\bjtitle{Science}
\bvolume{386},
\bfpage{76}.
\doiurl{https://doi.org/10.1126/science.ado2993}.
\adsurl{2024Sci...386...76Y}.
\end{barticle}
\endbibitem

\bibitem[\protect\citeauthoryear{{Yokoyama} and {Shibata}}{1996}]{yoko96}
\begin{barticle}
\bauthor{\bsnm{{Yokoyama}}, \binits{T.}},
\bauthor{\bsnm{{Shibata}}, \binits{K.}}:
\byear{1996},
\batitle{{Numerical Simulation of Solar Coronal X-Ray Jets Based on the
  Magnetic Reconnection Model}}.
\bjtitle{\pasj}
\bvolume{48},
\bfpage{353}.
\doiurl{https://doi.org/10.1093/pasj/48.2.353}.
\adsurl{1996PASJ...48..353Y}.
\end{barticle}
\endbibitem

\bibitem[\protect\citeauthoryear{{Zhang}}{2021}]{zhang21}
\begin{barticle}
\bauthor{\bsnm{{Zhang}}, \binits{Q.M.}}:
\byear{2021},
\batitle{{A revised cone model and its application to non-radial prominence
  eruptions}}.
\bjtitle{\aap}
\bvolume{653},
\bfpage{L2}.
\doiurl{https://doi.org/10.1051/0004-6361/202141982}.
\adsurl{2021A&A...653L...2Z}.
\end{barticle}
\endbibitem

\bibitem[\protect\citeauthoryear{{Zhang}}{2022}]{zhang22}
\begin{barticle}
\bauthor{\bsnm{{Zhang}}, \binits{Q.M.}}:
\byear{2022},
\batitle{{Tracking the 3D evolution of a halo coronal mass ejection using the
  revised cone model}}.
\bjtitle{\aap}
\bvolume{660},
\bfpage{A144}.
\doiurl{https://doi.org/10.1051/0004-6361/202142942}.
\adsurl{2022A&A...660A.144Z}.
\end{barticle}
\endbibitem

\bibitem[\protect\citeauthoryear{{Zhang} and {Ji}}{2014a}]{zqm14a}
\begin{barticle}
\bauthor{\bsnm{{Zhang}}, \binits{Q.M.}},
\bauthor{\bsnm{{Ji}}, \binits{H.S.}}:
\byear{2014}a,
\batitle{{A swirling flare-related EUV jet}}.
\bjtitle{\aap}
\bvolume{561},
\bfpage{A134}.
\doiurl{https://doi.org/10.1051/0004-6361/201322616}.
\adsurl{2014A&A...561A.134Z}.
\end{barticle}
\endbibitem

\bibitem[\protect\citeauthoryear{{Zhang} and {Ji}}{2014b}]{zqm14b}
\begin{barticle}
\bauthor{\bsnm{{Zhang}}, \binits{Q.M.}},
\bauthor{\bsnm{{Ji}}, \binits{H.S.}}:
\byear{2014}b,
\batitle{{Blobs in recurring extreme-ultraviolet jets}}.
\bjtitle{\aap}
\bvolume{567},
\bfpage{A11}.
\doiurl{https://doi.org/10.1051/0004-6361/201423698}.
\adsurl{2014A&A...567A..11Z}.
\end{barticle}
\endbibitem

\bibitem[\protect\citeauthoryear{{Zhang}, {Li}, and {Ning}}{2017}]{zqm17}
\begin{barticle}
\bauthor{\bsnm{{Zhang}}, \binits{Q.M.}},
\bauthor{\bsnm{{Li}}, \binits{D.}},
\bauthor{\bsnm{{Ning}}, \binits{Z.J.}}:
\byear{2017},
\batitle{{Simultaneous Transverse and Longitudinal Oscillations in a Quiescent
  Prominence Triggered by a Coronal Jet}}.
\bjtitle{\apj}
\bvolume{851},
\bfpage{47}.
\doiurl{https://doi.org/10.3847/1538-4357/aa9898}.
\adsurl{2017ApJ...851...47Z}.
\end{barticle}
\endbibitem

\bibitem[\protect\citeauthoryear{{Zhang} et~al.}{2012}]{zqm12}
\begin{barticle}
\bauthor{\bsnm{{Zhang}}, \binits{Q.M.}},
\bauthor{\bsnm{{Chen}}, \binits{P.F.}},
\bauthor{\bsnm{{Guo}}, \binits{Y.}},
\bauthor{\bsnm{{Fang}}, \binits{C.}},
\bauthor{\bsnm{{Ding}}, \binits{M.D.}}:
\byear{2012},
\batitle{{Two Types of Magnetic Reconnection in Coronal Bright Points and the
  Corresponding Magnetic Configuration}}.
\bjtitle{\apj}
\bvolume{746},
\bfpage{19}.
\doiurl{https://doi.org/10.1088/0004-637X/746/1/19}.
\adsurl{2012ApJ...746...19Z}.
\end{barticle}
\endbibitem

\bibitem[\protect\citeauthoryear{{Zhang} et~al.}{2020}]{zqm20}
\begin{barticle}
\bauthor{\bsnm{{Zhang}}, \binits{Q.M.}},
\bauthor{\bsnm{{Dai}}, \binits{J.}},
\bauthor{\bsnm{{Xu}}, \binits{Z.}},
\bauthor{\bsnm{{Li}}, \binits{D.}},
\bauthor{\bsnm{{Lu}}, \binits{L.}},
\bauthor{\bsnm{{Tam}}, \binits{K.V.}},
\bauthor{\bsnm{{Xu}}, \binits{A.A.}}:
\byear{2020},
\batitle{{Transverse coronal loop oscillations excited by homologous
  circular-ribbon flares}}.
\bjtitle{\aap}
\bvolume{638},
\bfpage{A32}.
\doiurl{https://doi.org/10.1051/0004-6361/202038233}.
\adsurl{2020A&A...638A..32Z}.
\end{barticle}
\endbibitem

\bibitem[\protect\citeauthoryear{{Zhang} et~al.}{2021}]{zqm21}
\begin{barticle}
\bauthor{\bsnm{{Zhang}}, \binits{Q.M.}},
\bauthor{\bsnm{{Huang}}, \binits{Z.H.}},
\bauthor{\bsnm{{Hou}}, \binits{Y.J.}},
\bauthor{\bsnm{{Li}}, \binits{D.}},
\bauthor{\bsnm{{Ning}}, \binits{Z.J.}},
\bauthor{\bsnm{{Wu}}, \binits{Z.}}:
\byear{2021},
\batitle{{Spectroscopic observations of a flare-related coronal jet}}.
\bjtitle{\aap}
\bvolume{647},
\bfpage{A113}.
\doiurl{https://doi.org/10.1051/0004-6361/202038924}.
\adsurl{2021A&A...647A.113Z}.
\end{barticle}
\endbibitem

\bibitem[\protect\citeauthoryear{{Zhang} et~al.}{2022a}]{zqm22a}
\begin{barticle}
\bauthor{\bsnm{{Zhang}}, \binits{Q.M.}},
\bauthor{\bsnm{{Chen}}, \binits{J.L.}},
\bauthor{\bsnm{{Li}}, \binits{S.T.}},
\bauthor{\bsnm{{Lu}}, \binits{L.}},
\bauthor{\bsnm{{Li}}, \binits{D.}}:
\byear{2022}a,
\batitle{{Transverse Coronal-Loop Oscillations Induced by the Non-radial
  Eruption of a Magnetic Flux Rope}}.
\bjtitle{\solphys}
\bvolume{297},
\bfpage{18}.
\doiurl{https://doi.org/10.1007/s11207-022-01952-3}.
\adsurl{2022SoPh..297...18Z}.
\end{barticle}
\endbibitem

\bibitem[\protect\citeauthoryear{{Zhang} et~al.}{2024}]{zqm24}
\begin{barticle}
\bauthor{\bsnm{{Zhang}}, \binits{Q.M.}},
\bauthor{\bsnm{{Lin}}, \binits{M.S.}},
\bauthor{\bsnm{{Yan}}, \binits{X.L.}},
\bauthor{\bsnm{{Dai}}, \binits{J.}},
\bauthor{\bsnm{{Hou}}, \binits{Z.Y.}},
\bauthor{\bsnm{{Li}}, \binits{Y.}},
\bauthor{\bsnm{{Qiu}}, \binits{Y.}}:
\byear{2024},
\batitle{{Two successive EUV waves and a transverse oscillation of a quiescent
  prominence}}.
\bjtitle{\mnras}
\bvolume{533},
\bfpage{3255}.
\doiurl{https://doi.org/10.1093/mnras/stae1936}.
\adsurl{2024MNRAS.533.3255Z}.
\end{barticle}
\endbibitem

\bibitem[\protect\citeauthoryear{{Zhang} et~al.}{2022b}]{zqm22b}
\begin{barticle}
\bauthor{\bsnm{{Zhang}}, \binits{Q.}},
\bauthor{\bsnm{{Li}}, \binits{C.}},
\bauthor{\bsnm{{Li}}, \binits{D.}},
\bauthor{\bsnm{{Qiu}}, \binits{Y.}},
\bauthor{\bsnm{{Zhang}}, \binits{Y.}},
\bauthor{\bsnm{{Ni}}, \binits{Y.}}:
\byear{2022}b,
\batitle{{First Detection of Transverse Vertical Oscillation during the
  Expansion of Coronal Loops}}.
\bjtitle{\apjl}
\bvolume{937},
\bfpage{L21}.
\doiurl{https://doi.org/10.3847/2041-8213/ac8e01}.
\adsurl{2022ApJ...937L..21Z}.
\end{barticle}
\endbibitem

\bibitem[\protect\citeauthoryear{{Zhang} et~al.}{2023}]{zqm23}
\begin{barticle}
\bauthor{\bsnm{{Zhang}}, \binits{Q.}},
\bauthor{\bsnm{{Zhou}}, \binits{Y.}},
\bauthor{\bsnm{{Li}}, \binits{C.}},
\bauthor{\bsnm{{Li}}, \binits{Q.}},
\bauthor{\bsnm{{Xia}}, \binits{F.}},
\bauthor{\bsnm{{Qiu}}, \binits{Y.}},
\bauthor{\bsnm{{Dai}}, \binits{J.}},
\bauthor{\bsnm{{Zhang}}, \binits{Y.}}:
\byear{2023},
\batitle{{Transverse Vertical Oscillations During the Contraction and Expansion
  of Coronal Loops}}.
\bjtitle{\apj}
\bvolume{951},
\bfpage{126}.
\doiurl{https://doi.org/10.3847/1538-4357/acd5cf}.
\adsurl{2023ApJ...951..126Z}.
\end{barticle}
\endbibitem

\bibitem[\protect\citeauthoryear{{Zhong} et~al.}{2023}]{zhong23}
\begin{barticle}
\bauthor{\bsnm{{Zhong}}, \binits{S.}},
\bauthor{\bsnm{{Nakariakov}}, \binits{V.M.}},
\bauthor{\bsnm{{Kolotkov}}, \binits{D.Y.}},
\bauthor{\bsnm{{Chitta}}, \binits{L.P.}},
\bauthor{\bsnm{{Antolin}}, \binits{P.}},
\bauthor{\bsnm{{Verbeeck}}, \binits{C.}},
\bauthor{\bsnm{{Berghmans}}, \binits{D.}}:
\byear{2023},
\batitle{{Polarisation of decayless kink oscillations of solar coronal loops}}.
\bjtitle{Nature Communications}
\bvolume{14},
\bfpage{5298}.
\doiurl{https://doi.org/10.1038/s41467-023-41029-8}.
\adsurl{2023NatCo..14.5298Z}.
\end{barticle}
\endbibitem

\bibitem[\protect\citeauthoryear{{Zhou} et~al.}{2024}]{zhou24}
\begin{barticle}
\bauthor{\bsnm{{Zhou}}, \binits{X.}},
\bauthor{\bsnm{{Tang}}, \binits{Z.}},
\bauthor{\bsnm{{Qu}}, \binits{Z.}},
\bauthor{\bsnm{{Yu}}, \binits{K.}},
\bauthor{\bsnm{{Zhou}}, \binits{C.}},
\bauthor{\bsnm{{Xiang}}, \binits{Y.}},
\bauthor{\bsnm{{Ibrahim}}, \binits{A.A.}},
\bauthor{\bsnm{{Shen}}, \binits{Y.}}:
\byear{2024},
\batitle{{On the Origin of a Broad Quasiperiodic Fast-propagating Wave Train:
  Unwinding Jet as the Driver}}.
\bjtitle{\apjl}
\bvolume{974},
\bfpage{L3}.
\doiurl{https://doi.org/10.3847/2041-8213/ad7a68}.
\adsurl{2024ApJ...974L...3Z}.
\end{barticle}
\endbibitem

\bibitem[\protect\citeauthoryear{{Zimovets} and {Nakariakov}}{2015}]{zim15}
\begin{barticle}
\bauthor{\bsnm{{Zimovets}}, \binits{I.V.}},
\bauthor{\bsnm{{Nakariakov}}, \binits{V.M.}}:
\byear{2015},
\batitle{{Excitation of kink oscillations of coronal loops: statistical
  study}}.
\bjtitle{\aap}
\bvolume{577},
\bfpage{A4}.
\doiurl{https://doi.org/10.1051/0004-6361/201424960}.
\adsurl{2015A&A...577A...4Z}.
\end{barticle}
\endbibitem

\bibitem[\protect\citeauthoryear{{Zimovets} et~al.}{2021}]{zim21}
\begin{barticle}
\bauthor{\bsnm{{Zimovets}}, \binits{I.V.}},
\bauthor{\bsnm{{McLaughlin}}, \binits{J.A.}},
\bauthor{\bsnm{{Srivastava}}, \binits{A.K.}},
\bauthor{\bsnm{{Kolotkov}}, \binits{D.Y.}},
\bauthor{\bsnm{{Kuznetsov}}, \binits{A.A.}},
\bauthor{\bsnm{{Kupriyanova}}, \binits{E.G.}},
\bauthor{\bsnm{{Cho}}, \binits{I.-H.}},
\bauthor{\bsnm{{Inglis}}, \binits{A.R.}},
\bauthor{\bsnm{{Reale}}, \binits{F.}},
\bauthor{\bsnm{{Pascoe}}, \binits{D.J.}},
\bauthor{\bsnm{{Tian}}, \binits{H.}},
\bauthor{\bsnm{{Yuan}}, \binits{D.}},
\bauthor{\bsnm{{Li}}, \binits{D.}},
\bauthor{\bsnm{{Zhang}}, \binits{Q.M.}}:
\byear{2021},
\batitle{{Quasi-Periodic Pulsations in Solar and Stellar Flares: A Review of
  Underpinning Physical Mechanisms and Their Predicted Observational
  Signatures}}.
\bjtitle{\ssr}
\bvolume{217},
\bfpage{66}.
\doiurl{https://doi.org/10.1007/s11214-021-00840-9}.
\adsurl{2021SSRv..217...66Z}.
\end{barticle}
\endbibitem

\end{thebibliography}

\end{article}
\end{document}